\documentclass[prapplied,aps,10pt,twocolumn,showpacs,amsmath,amssymb,superscriptaddress,floatfix,longbibliography,nofootinbib]{revtex4-2}

\usepackage{times}
\usepackage{helvet}
\usepackage{courier}
\usepackage{hyperref}
\hypersetup{
    colorlinks=true,       
    linkcolor=cyan,          
    citecolor=magenta,        
    filecolor=magenta,      
    urlcolor=RoyalBlue,           
    runcolor=cyan
}

\usepackage{graphicx}
\usepackage{natbib}
\usepackage{bibunits}

\usepackage{newfloat}
\usepackage{listings}

\usepackage{amsmath}
\usepackage{amsfonts}
\usepackage{amssymb}
\usepackage{amsthm}
\usepackage{braket}
\usepackage[capitalize]{cleveref}
\usepackage[dvipsnames]{xcolor}
\usepackage{booktabs}

\usepackage{mathtools}
\usepackage{dsfont}

\begin{document}

\title{Applications and resource estimates for open system simulation on a quantum computer}

\author{Evgeny Mozgunov}\thanks{mvjenia@gmail.com}
\affiliation{Center for Quantum Information Science \& Technology, University of Southern California}
\affiliation{Department of Electrical \& Computer Engineering, University of Southern California}

\author{Jeffrey Marshall}
\affiliation{Quantum Artificial Intelligence Laboratory, NASA Ames Research Center, Moffett Field, CA 94035, USA}
\affiliation{USRA Research Institute for Advanced Computer Science, Mountain View, CA 94043, USA}

\author{Namit Anand}
\affiliation{Quantum Artificial Intelligence Laboratory, NASA Ames Research Center, Moffett Field, CA 94035, USA}
\affiliation{KBR, Inc., 601 Jefferson St., Houston, TX 77002, USA}

\begin{abstract}
We present two applications where open system quantum simulation is the preferred approach on a quantum computer. We choose concrete parameters for the problems in such a way that the application value, which we call utility, can be obtained from the solution directly. The scientific utility is exemplified by a computation of nonequilibrium behavior of Ca$_3$Co$_2$O$_6$, which is studied in \$2M MagLab experiments. For industrial utility, we develop a methodology that allows researchers of various backgrounds to estimate the economic value of an emerging technology consistently. Our approach predicts \$400M utility for the applications of materials with a Metal-Insulator Transition. We focus on the transport calculation in the Hubbard model as the simplest problem that needs to be solved in a large-scale material search. The resource estimates for both problems suffer from a large required runtime, which motivates us to propose novel algorithm optimizations, taking advantage of the translation invariance and the parallelism of the T-gate application. Finally, we introduce several planted solution problems and their obfuscated versions as a benchmark for future quantum devices. 

\end{abstract}

\maketitle

\tableofcontents

\section{Introduction}
We study applications of an open system quantum simulator. Such a simulator can be realized by a universal quantum computer \cite{cleve2016efficient}, and some classical methods can approximate it as well. Our goal is to introduce a concrete computational problem such that finding its solution would have direct scientific or industrial utility. We find an example of the scientific utility to be the investigation of materials that are already studied in high-cost experiments. 20-30 new material samples per year are not understood by low-cost experimental and computational means and are sent to the National High Magnetic Field Laboratory (MagLab) \cite{national2013high}. These experiments often observe unexplained nonequilibrium phenomena, exemplified by the behavior of Ca$_3$Co$_2$O$_6$ \cite{nekrashevich2022reaching}. We use them as application instances. 

For industrial utility, we focus on material searches. A successful example of a large throughput material search is the search for material for good thermoelectric devices, happening under the umbrella of the Materials Project \cite{jain2013commentary}. There are already commercial applications for those devices, but wide adoption is still dependent on the discovery of better materials and designs using the semi-automated search methodology. The utility of this computational search comes from accelerating their adoption and the projected big markets for applications. Unfortunately for quantum computers, we believe the bottlenecks in this search are largely classical, and most questions can and should be answered using cheaper state-of-the-art classical algorithms. Drawing inspiration from the success of these classical techniques, we envision a similar search in the domain of strongly correlated materials. There, classical methods fail, though there are some promising new results \cite{cui2023ab}. The phenomena that are most often discussed in the context of applications are high-temperature superconductivity and the metal-insulator transition (MIT). The latter is especially suited for an open system study, as one of the central observables is the transport coefficient. It can, in principle, be estimated in a closed system setting, yet we believe the most efficient quantum algorithm for transport contains an open system quantum simulator. The corresponding economic impact will be outlined below. The majority of the MIT search will need to investigate ab initio models similar to those considered in \cite{cui2023ab}; however, as a starting point, we choose a simpler Hubbard model as a lower bound of the required computational resources of this approach.

We also note that many other application instances covered in companion publications \cite{watts2024fullereneencapsulatedcyclicozonegeneration,otten2024quantumresourcesrequiredbinding,nguyen2024quantumcomputingcorrosionresistantmaterials,bellonzi2024feasibilityacceleratinghomogeneouscatalyst,penuel2024feasibilityacceleratingincompressiblecomputational,elenewski2024prospectsnmrspectralprediction,agrawal2024quantifyingfaulttolerantsimulation,bartschi2024potentialapplicationsquantumcomputing,saadatmand2024faulttolerantresourceestimationusing}, while not directly related to open systems, are expected to be solved by quantum algorithms that rely on an unproven assumption of the existence of a high overlap initial state. Estimating that overlap is a challenge, and here we note that for open system thermal state preparation relevant to most condensed matter and material science applications, as well as in quantum chemistry such as surface chemistry, the estimate can be performed under very mild assumptions compared to other state preparation methods.

In this work, we perform resource estimates for quantum algorithms solving our chosen open system simulation problems for Ca$_3$Co$_2$O$_6$ and the Hubbard model. The complexity of the best open system quantum simulation algorithms is only known in terms of big-O notation, which limits our estimates as we ignore the prefactors. Prefactors can be readily estimated for suboptimal quantum simulation. While simulating the high-temperature Hubbard model can be performed within a realistic time scale of hours, low-temperature ab initio simulations appear unfeasible without further algorithmic improvements. An even simpler complication we encounter is that for any material like Ca$_3$Co$_2$O$_6$, where nonequilibrium dynamics takes weeks of real time, the ratio between the microscopic time scale and the timescale of the slow dynamics is $10^{11}$ or higher. This ratio directly multiplies the required algorithm runtime, resulting in unrealistic numbers of the order of a billion years. The only ways around this issue are fast-forwarding of the dynamics or utilizing NISQ quantum simulators such as D-wave \cite{king2021quantum}. As the runtime appears to be the deciding factor in the practicality of these algorithms, we adopt an unusual hardware compilation scheme. The conventional approach \cite{beverland2022assessing,harrigan2024expressing} gets rid of all the Clifford operations, which forces it to perform remaining T-gates sequentially. Our estimates and algorithm optimizations utilizing translation invariance take advantage of the parallel application of T-gates. This can be realized in hardware if Clifford gates are performed explicitly. The CNOTs that are not spatially local, in particular, need to be performed via logical qubit teleportation, where entangled Bell pairs become a resource consumed throughout the computer in a similar way to magic states. This approach has a high physical qubit count, which is the tradeoff for better runtime.

In Sec. \ref{sec:ut} we discuss our general methodology of utility estimation and apply it to cases of interest. The relevant economic concepts are further discussed in  App. \ref{app:eco}. Then, in Sec \ref{sec:ai}, we state two of the cases of interest as fully specified mathematical problems. We list available quantum algorithms in Sec. \ref{sec:qi} before describing the implementation of the best of them for our problems in Sec. \ref{sec:re}, calculating zeroth order estimates for both logical and physical required resources. In App. \ref{sec:app} we use toy problem benchmarks where we can compute the answer to test the capabilities of the future quantum device. The intended use of those planted solution benchmarks is to match the size of the application benchmarks so that solving one guarantees the ability to solve the other. App. \ref{app:howtocount} describes how to count T-gates and depth for the operations used in our algorithm and App. \ref{app:av1019} provides estimates for a concrete computational problem that is a proxy for the average case of the MIT material search (in contrast to Hubbard, which was the easiest case). App. \ref{app:plqtr} describes how we utilized the \textsc{pyLIQTR} \cite{pyLIQTR} library for the block encoding of the Hubbard model Hamiltonian, and in App. \ref{app:ex} we compute the exact resource estimates for a simpler suboptimal algorithm to contrast it with the order of magnitude estimates used in the main text.

\section{Utility}
\label{sec:ut}
There are two approaches to the utility of scientific findings: to look into the funding that science receives and to look into the market of inventions enabled by the discovery in question. The latter is possible only in hindsight, while the former may be insensitive to differences between fundamental and applied research. For private research, one may assume an efficient market, which would require both estimates to result in a similar order of magnitude. The efficiency of government-funded research depends on the policy of the corresponding funding agency. We expect it to be at least within an order of magnitude of efficiency. 

We note that the quantum computer is not necessarily the only way to accomplish the research agenda we outline. Whether it would be accomplished by classical computation, experiment, quantum computer, or a combination of those approaches, the utility is unchanged. In particular, the presence or absence of quantum advantage should not affect our estimate of utility. Nonetheless, we focus on the problems that promise a quantum advantage as well. In the following, we will outline the methodology of each approach to utility estimation and try both approaches for the chosen application instances.

\subsection{Scientific utility: funding}

For the funding estimate, we look into the amount of money spent on solving the problem in question or a problem of equivalent size and significance. The rationale for that is that one capable of solving the problem should be able to negotiate for that amount of remuneration, at least in principle. We note that a transaction like that would not necessarily occur. For example, the creators of AlphaFold \cite{Alphafold2021} did not monetize it in a way commensurate with the amount of money regularly spent on protein structure experiments and simulations. Possibly, the publicity it generated was of equivalent value to them, and it undoubtedly accelerated biological research, so we believe estimating its value is still appropriate using our methodology even if the hypothetical transaction did not take place.

There is a subtle difference between estimating the utility of an answer to one fully specified computational problem, and the utility of a black box solving any problem in some general class. We focus on the former, while \cite{agrawal2024quantifyingfaulttolerantsimulation} establish a data-driven methodology for the latter, using the publication data in the corresponding field.

\subsection{General methodology for industrial utility}
The argument that a particular technology's market size is the utility of further research into it is very common. Almost any research is conventionally justified in that way, no matter how distant it is from applications. We would like to challenge this way of thinking by proposing a methodology that would result in a consistent (within 1-2 orders of magnitude) estimate of utility even when applied by different researchers and decision-makers independently. While it is true that it is sometimes impossible to trace the influences of fundamental research on applications far in the future, most of the revolutionary technologies of the past have had clear conditions for their discovery. We choose to focus on concrete links between research and applications. Thus, only the research that is part of a clearly defined roadmap to an improvement in the corresponding technology will be assigned nonzero utility in this approach. In this way, we will estimate the effort required for the successful completion of that roadmap and assign the utility to that. 

To apply our proposed methodology to quickly and consistently estimate the industrial utility of a particular research program, three pieces of information about it are required:

\begin{enumerate}
\item A clear roadmap incorporating the solution of the research problem into the design of the product
\item A scientific review listing competing technologies for this type of product
\item A market report on the size of the market (sum of yearly revenue) of competing products
\end{enumerate}
We will now describe how to utilize these sources.

\paragraph{Scientific review} The risk associated with new technology is that it will not reach the specifications required for mass adoption and, instead, be outmatched by competitors relying on different technologies. In this way, various research directions race against each other in terms of performance metrics. Such performance metrics for different technologies can usually be found in a contemporary scientific review. At the early stages of the race, we choose to divide the utility estimate in a way weighted by the performance metrics, such that the most promising candidates correspond to the larger share of utility. Since the performance metrics vary across fields, and both logarithms and the value itself are being used, we need to set up a unified way to assign those weights. The simplest way is to order $n$ competitive technologies by how promising they are and assign weights $\sim n, n-1 \dots 1$ to them. The utility of each technology will then be $=$weight$\cdot$utility of the application. 

Our approach may fail in more complex situations such as when one of the technologies has already received mass adoption. In that case, while there is a probability that the entire industry will switch to a new technology, complete switches like that are rare. E.g., no one is building coal plants in the US anymore, but the existing ones are still in use, and sometimes their operation is even extended beyond the planned closure date \cite{coal}. The switch may not be complete and will have a lot of inertia. The available data does not allow us to reliably estimate this switch probability - indeed, it is hard to formulate a methodology that will result in the same (within 1-2 orders of magnitude) estimate when applied independently by different analysts or research groups. Instead, we rely on a more tangible picture that the new technology will occupy a niche of the market where it is especially advantageous. 

\paragraph{Market report} The data for such market segmentation is usually available in the industry reports published by consulting companies. Those reports can be rather controversial, such as the trillion-dollar estimates of the utility of quantum computers themselves \cite{bogobowicz2024steady}. Below, we outline the methodology we use to interpret those reports, as well as the precautions that need to be taken when using them.

The market size of a technology is commonly estimated as the sum of the yearly revenue of the companies utilizing that technology. More specifically, if a small part of a big company is using it, the corresponding share is used. The revenue is the product of the product's price and the number of sales. For simplicity, we will assign a factor between $10$ (technology will lead to an explosive growth of the business) and $1/10$ depending on how central the technology is for the business model. We discuss the possible considerations going into this number further in App. \ref{app:eco}. For example, if energy is the leading cost and improvement in efficiency directly translates to energy savings, such as for nitrogen fixation, it is reasonable to choose the factor $1/2$. For thermoelectric devices, the factor is $10$: an improvement in efficiency and other qualities may dramatically increase the market, unlocking the main application of energy recycling.

A related question is, at what time should we take the revenue? With the quantum computer application in mind, we need to wait for at least ten years before the solution to the problem becomes available. Most projections of the market size stop at the 10-year time horizon. Thus, we will use an estimate of the market size ten years in the future (year 2034) corresponding to the emerging technology or competing technology in case the market is already occupied by existing technology. Unfortunately, the market breakdown based on emerging technologies is not commonly found in consulting reports. At best, we can narrow it down to the niche of the existing technology to be replaced. What is common is  another type of estimate: instead of looking at the revenue of the seller of the technology, a report would look at the revenue of all the potential buyers of the technology. Then, the report would note that an improvement in the product would generate some arbitrary fraction, such as a 1\%-9\% increase in the revenue of the buyers. The possibility of such an improvement is confirmed by experts in the buyers' corresponding businesses. Thus, the total value is computed as a sum of 1\%-9\% of the revenues of all buyers, and the resulting number for some of the future technologies, such as AI or quantum computing, is quoted in the trillions \cite{bogobowicz2024steady}. The apparent discrepancy with the \$600b total semiconductor market is intentional: the message of those reports is that the semiconductor companies could generate more revenue, pushing the software middlemen out and realizing their full potential. This message is likely biased: it is a sales pitch for the semiconductor companies to hire the consulting services of the company that made the report.  For our purposes, we will only look at the value the seller companies were able to realize as revenue, not the optimistic possibility that is quoted in the more controversial reports. Moreover, only when a company based in one country can, in principle, take over the entire global market with a superior technology that cannot be reverse-engineered can we use global revenues instead of the revenue of companies based in one particular country. 

A final caution is that fake consulting reports exist. Especially for emerging technologies such as neuromorphic computing, all the reports available online are of dubious origin and quote seemingly random numbers as the market size. A well-known consulting company's reports are available for the dedicated AI hardware but not for the neuromorphic devices. The companies invested in them likely have not generated enough revenue to report. Assuming that the dedicated AI hardware of the future will use neuromorphic devices, we can use the dedicated AI hardware market size and estimate the probability of the switch in the way outlined above. The estimate of the market size of a segment like that may vary by two orders of magnitude between reports. Whenever possible, we average the reports of well-known consulting companies, thus achieving one order of magnitude error. As the other factors we used are ad hoc, this estimate would result in two orders of magnitude error. The formula is as follows:
\begin{align}
   \text{Utility} = \text{weight of tech.} ~ \times \text{share in revenue} ~\times \\ \times  ~\text{global market size of niche tech. }
\end{align}

We expect the two utility estimates (science funding and market size) to match within the two orders of magnitude error.

\subsection{MagLab single material \$ cost}
MagLab pipeline contains costly experiments with high magnetic fields. The cost of that pipeline is known (\$39M a year is the budget of MagLab), and it is also known that classical algorithms do not fully explain Ca$_3$Co$_2$O$_6$ experiment \cite{nekrashevich2022reaching} proposed as a benchmark. The open system contribution to the dynamics of that benchmark is likely to be important for reproducing the experimental behavior and eventually replacing the experiment altogether. 

In our case, the utility would be the MagLab budget divided by the number of materials going through it, thus about \$2M per material.

The specific experiment \cite{nekrashevich2022reaching} concerning the nonequilibrium spin states does not have direct applications. In general, spin materials are studied for the advancement of:
\begin{itemize}
    \item Next-gen storage and computing (especially QC hardware)
    \item Magnetoelectrics, piezomagnets, sensors
    \item Superconductors
\end{itemize}

While Ca$_3$Co$_2$O$_6$ itself is used in prototypes of fuel cells at room temperature and zero magnetic field \cite{wei2013evaluation}, a concrete application for its low-temperature high field behavior has not been invented. We conclude the market size estimate at \$0 according to our methodology.

\subsection{Metal-Insulator transition material search \$ cost}
In this section, we discuss the utility of a material search for new materials with Metal-Insulator Transition (MIT). First of all, we note that this transition happens at room temperature, so it may be possible to predict it ab initio classically, but existing attempts were not very successful. Our utility estimate will be valid either way, but it does not have to be a quantum computer that succeeds in this material search. A table of possible applications of MIT can be found in the supplement of \cite{lee2024nanoelectronics}, including:
\begin{itemize}
    \item Memory
    \item Logic
    \item Artificial Synapses
    \item Artificial Neurons
    \item Photodetectors 
    \item Gas Sensors
\end{itemize}

 While most research focuses on making a transistor out of MIT material, it is not expected that it would replace conventional semiconductor technology for integrated circuits. Indeed,  a literature search in the reviews discussing the continuation of Moore's law showed that MIT materials are not listed as potential candidates for a new transistor for conventional computers. There is a field, however, where it is considered promising: the field of neuromorphic computing. Indeed, the reviews \cite{tang2019bridging} would list it as the 3rd best out of 6 potential candidate technologies, giving the weight factor of 1/5. There is no direct estimate for the market size of neuromorphic computing, but we could use the entire AI market share, which is 20\% of the semiconducting hardware market as listed in \cite{batra2019artificial}. The US semiconductor market is \$80b, while the global one is \$600b. Ten years from now, it is projected to grow to \$1,100b. We do not expect that the chip production using a completely new paradigm of neuromorphic computing can easily be reverse-engineered over a 10-year timescale, assuming that the details of the fabrication process are kept private early on, which means we can use the global revenue for our estimate.  As the main financial investment in semiconductor manufacturing is the expensive fabrication facilities, not the research into the transistor technology itself, we estimate a 1/10 ratio of how the chip technology will factor in the revenue. Finally, we need to estimate the probability of switching from the current dedicated AI hardware to neuromorphic computing. Again, since there are various competing technologies (GPU, TPU, etc.), we estimate another 1/10 ratio as the neuromorphic devices take the last place out of 4 candidates. The resulting utility estimate is:
\begin{equation}
    \text{Utility} = \frac{1}{5}\times \frac{1}{5} \times \frac{1}{10}\times \frac{1}{10}\times \$1,100\text{b} = \$440\text{M}
\end{equation}

The size of the research grants spent on this search so far can be estimated from the reports on DOE programs and European collaborations. Q-MEEN-C in San Diego has been funded for eight years with \$3M per year, focusing on MIT materials and spintronics for neuromorphic computing. It also contains experimental effort in fabricating device prototypes. We expect roughly 1/10 of funding to go to the computational efforts, resulting in a \$2.4M scientific funding estimate of utility. We also note the Materials Project, from which we learn that about 100 materials with MIT are known \cite{georgescu2021database}, though the funding structure of that project does not allow an easy estimate, and 1.5M euro FIRSTORM program as an example of a smaller scale computational research program on this topic. An automated estimate of the total size of funded computational research programs on this topic can be performed following the methodology of \cite{agrawal2024quantifyingfaulttolerantsimulation}. We leave it for future work, assuming the result would come out to \$40M in the programs funded today. As only 100 materials were found, and for the full search, we assume that 1000 materials need to be studied, we expect that the total cost would match the $\sim$\$400M industry estimate. 

\subsection{Gibbs state preparation: most likely targets}

In this section, we discuss a subroutine of many quantum algorithms: using a thermal state as an initial state. Applications that allow for those initial states are the material searches such as the high-Tc and MIT material searches, as well as quantum chemistry problems that involve modeling materials such as surface chemistry. Some problems in fundamental science also require thermal initial states, such as the MagLab experiment on Ca$_3$Co$_2$O$_6$ described above. We estimate the total utility of all the applications where Gibbs state preparation is a reliable tool to be of the order of \$100M and refer to companion publications \cite{watts2024fullereneencapsulatedcyclicozonegeneration,otten2024quantumresourcesrequiredbinding,nguyen2024quantumcomputingcorrosionresistantmaterials,bellonzi2024feasibilityacceleratinghomogeneouscatalyst,penuel2024feasibilityacceleratingincompressiblecomputational,elenewski2024prospectsnmrspectralprediction,agrawal2024quantifyingfaulttolerantsimulation,bartschi2024potentialapplicationsquantumcomputing,saadatmand2024faulttolerantresourceestimationusing} for specific resource and utility estimates. While Gibbs state preparation alone does not solve those applications, and each individual task in those \$100M can be only \$2M or less, with varying required number of runs, it may well be possible that Gibbs state preparation will dominate the resource cost. The reasons why one would still choose Gibbs state preparation instead of other approaches are explained below. 

We first digress to discuss a more common problem of ground state preparation. To obtain the desired precision, existing algorithms of ground state preparation usually contain an unknown factor in the dependence of the total resource cost on the system size. For example, a ground state estimation algorithm can enhance the probability of the ground state for an initial state with nonzero overlap, with circuit size growing with the inverse overlap. The initial state is then constructed by another circuit, and the cost of that circuit is impossible to compute before the quantum computer actually becomes available. It is sometimes assumed that this circuit's cost is negligible, so only the ground state estimation resource cost is counted in the resource estimate. It is, in general, not known what is the class of Hamiltonians where the ground state can be obtained to polynomial accuracy in polynomial time. 1D matrix product states (MPS) of gapped Hamiltonians are one example of such a system \cite{dalzell2019locally}: it is proven that the MPS can be computed classically and then constructed as a circuit inside the quantum computer. The overheads of such a construction may be quite big, though. Besides the rigorously proven results, we also have intuition about the behavior of the quantum annealing approach to the preparation of the polynomial overlap state. Essentially, a path avoiding first-order phase transitions needs to be found connecting the target state to some accessible initial state, such as the lower overlap state prepared by other methods or the product state. First-order phase transitions, however, are quite common. Thus, the adiabatic state preparation cannot be applied blindly. Every target system's phase diagram needs to be analyzed independently. We see that there is no unified method for providing a resource estimate for the state preparation of application-relevant states. The best one can hope for is an extrapolation for small system sizes, where both positive and negative results have been claimed \cite{lee2023evaluating} since the accuracy of the extrapolation is not controlled by anything. 

In comparison, thermal state preparation allows for more accessible guarantees. Indeed, the thermal state of any system can be prepared for sufficiently high temperatures \cite{alhambra2021locally}. Moreover, a bound exists \cite{chen2021fast} for the time to prepare the thermal state for all systems satisfying the Eigenstate Thermalization Hypothesis (ETH). Extrapolating the parameters in that bound from the exact diagonalization results would be more stable than extrapolating the overlaps in ground state estimation since those parameters reflect physical phenomena. The distinction between ETH and non-ETH systems is also well understood, with the latter containing many-body localization, spin glasses, and integrable models. Most of the application models are not in any of those categories. For example, glassy effects have never been observed in nonequilibrium electron dynamics of large molecules in quantum chemistry. That suggests that we could reliably prepare finite temperature states in those systems by a universal algorithm. Compare it with the adiabatic ground state preparation, which relied on avoiding first-order phase transitions above. To the best of our knowledge, thermal state preparation is the only reliable way to obtain resource counts for the initial state preparation algorithms.
\section{Problem Statement}
\label{sec:ai}
In both application instances, we will formulate an open system evolution given by the master equation:
\begin{equation}
    \frac{d\rho}{dt} =\mathcal{L}\rho := -i[H,\rho] + \sum_{\mu=1}^M (2L_\mu \rho L_\mu^\dag - \{L_\mu^\dag L_\mu, \rho\}) 
\end{equation}
by specifying the Hamiltonian $H$, initial state $\rho(0)$ and the Lindblad generators $L_\mu$. While most of our estimates use simple local forms of $L_\mu$, only a carefully picked $L_\mu$ has a correct thermal steady state. For instance, \cite{chen_quantum_2023} uses $L_\mu^{\text{therm}}$ defined via Hamiltonian evolution $e^{iHt}$ of a simple local operator. While such an operator would generally span the entire system and contain exponentially many Pauli's, it can be evaluated in polynomial time on a quantum computer. The prefactors of that approach, however, can be prohibitive, as we will show for the average case of MIT material search. We believe that a more practical approach would be to optimize $L_\mu$ on a classical computer such that it produces a thermal state approximately while maintaining a simple local form. In the third, broad-scope application instance of preparing Gibbs states, we expect dedicated Gibbs state preparation algorithms (also proposed in \cite{chen_quantum_2023}, as well as \cite{cubitt2023dissipative,chen2021fast}) to be more efficient than a simulation of an open system evolution. In all of these algorithms, the algorithm runtime necessary for equilibration is a property of the chosen approach and is not guaranteed to be polynomial except in special cases of high temperature or low dimensionality. In principle, it can be extrapolated from the sizes that can be addressed classically. If those sizes prove too small, one can use the formula in \cite{chen2021fast} relating the relaxation time to the parameters of the eigenstate thermalization hypothesis, which may be determined at smaller sizes. We have not investigated the necessary resource estimates further in this work.

\subsection{MagLab's application problem (Ca$_3$Co$_2$O$_6$)}
We choose this experiment as a target for quantum simulation because
Quantum-Classical separation can be argued for from the results of \cite{nekrashevich2022reaching}. Specifically, in Fig. 6 of that work, the results of Quantum Monte-Carlo (QMC) are shown as a proxy for experimental Fig. 7, showing nonequilibrium open system dynamics. Our goal is to reproduce Fig. 7 of \cite{nekrashevich2022reaching} via an {\textit{ab initio}} simulation. That is in contrast to QMC, which is a classical random process that only qualitatively reproduces the quantum nonequilibrium open system dynamics. In particular, the relationship between the number of spin updates and the time elapsed in the quantum dynamics, as well as the discretization of the imaginary time register, are fitting parameters when QMC is used as a proxy for nonequilibrium dynamics. They cannot be computed {\textit{ab initio}}. 

\begin{figure}[t]
\centering
\includegraphics[width=\columnwidth]{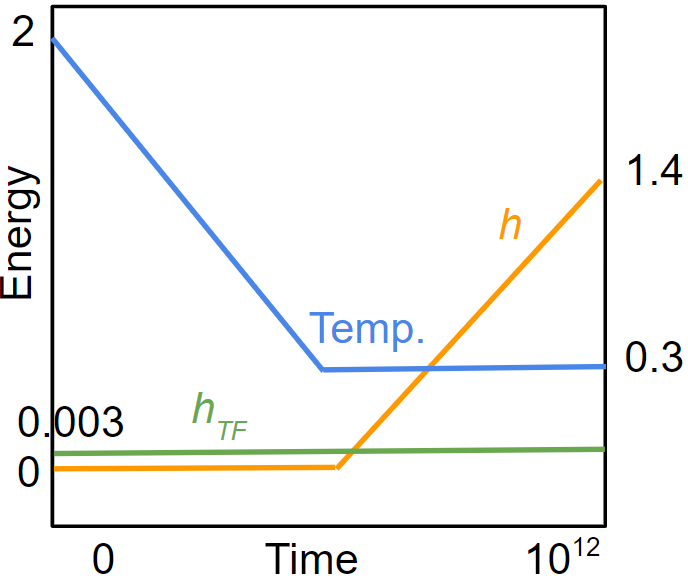}
\caption{The schedules used for a quantum simulation of an experiment \cite{nekrashevich2022reaching} shortened from 1 week to 1s, in energy units $|J_1|=1$ and inverse energy units for time}
\label{fig:sch}
\end{figure}

For the purposes of this experiment, Ca$_3$Co$_2$O$_6$ is well described by a version of the Transverse Field Ising model. The temperature and the longitudinal and transverse magnetic fields are controlled by the experimentalist, as shown in Fig. \ref{fig:sch}. We stress that while many MagLab studies correspond to closed system processes, the specific experiment shown in Fig. \ref{fig:sch} is a {\bf{nonequilibrium open system evolution}}. 

\begin{figure}[t]
\centering
\includegraphics[width=\columnwidth]{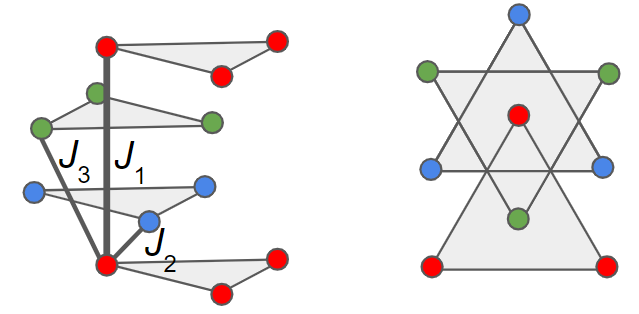}
\caption{Illustration of the layout of spins in Ca$_3$Co$_2$O$_6$. There are three alternating layers, each containing a triangular lattice of spins. The interactions are between layers, as shown. Four layers are shown on the left, and the view from the top is on the right.}
\label{fig:lay}
\end{figure}

The spins are arranged in three sublattices, as shown in Fig. \ref{fig:lay}. $J_1=-1$ is ferromagnetic, while $J_2=0.1|J_1|$ and $J_3=0.1|J_1|$ are antiferromagnetic. The overall energy scale is $23.9(2)$K, which can be translated into $4.98\cdot 10^{11}$Hz.  The z-component of the Hamiltonian is given by:
\begin{equation}
    H_z(t) = \sum_{\gamma=1,2,3}J_\gamma \sum_{(i,j)_\gamma} \sigma_i^z\sigma_j^z - h(t) \sum_i \sigma_i^z \label{CaH}
\end{equation}
An additional term is the transverse field $H_x =h_{TF} \sum_i \sigma_i^x$ with $h_{TF}=|J_1|/300$. The total Hamiltonian is $H(t) = H_z(t) + H_x$. The problem is to simulate the nonequilibrium states arising during the sweep of the longitudinal field $h(t)$. The longitudinal field $h(t)$ is swept from $0$ to $1.4|J_1|$ over a week of real time, $T=5.57\cdot 10^5$s $=2.77\cdot 10^{17}$ in the units where $|J_1|=1$ in the original experiment. We will only consider $T=1$s$=4.97\cdot10^{11}$ in natural units as it will turn out to be enough to illustrate how long nonequilibrium dynamics is still out of reach of quantum simulation. 

An open question is how large does a transversal field need to be ($h_{TF} \sim |J_1|$ or $2000$T is a conjecture) to remove the signatures of those nonequilibrium states - those fields cannot be probed in an experiment, and quantum simulation is the only tool that can access them. 

Here, we specify the open system addition to the above Hamiltonian that aims to reproduce the data shown in Fig. 7 of \cite{nekrashevich2022reaching}. The simplest way that may already describe the data is to add Lindblad operators $\sim \sigma^- =|0\rangle \langle 1|$ and $\sigma^+ =|1\rangle\langle0|$ conditioned on the states of neighboring spins, with rates given by the classical energies of the computational basis spin configurations. 
\begin{align}
 L_{i,r_\pm,\pm} = r_\pm P_{r_\pm}\sigma^\pm_i, \label{CaLi}\\    
 P_{r_\pm} =\sum_{s:r_{\pm,s} =r_\pm }\left(\prod_{j\in NN(i)} P_{j,s_j}\right)\\
 r_\pm \in \{ r_{\pm,s}\}, \quad r_{+,s}\ = \text{MC rates for }H_z   =\\
 =
 \begin{cases}
			e^{-\beta \Delta E}, & \text{if} \Delta E >0 \\
            1, & \text{otherwise}
		 \end{cases}
 \\
 \text{where }\Delta E =H_z^{(i)}(s,+1) -H_z^{(i)}(s,-1)   \ ,
\end{align}
and $P_{j,s_j} = \frac{1}{2} (\sigma^z_j + s_j)$ is a projector onto a state $|s_j\rangle$. The energies $H_z^{(i)}(s,s_0)$ are the terms of $H_z$ containing $\sigma^z_i$, after the substitution $\sigma^z_j \to s_j$ for all neighbors $j \in NN(i)$ in the interaction graph, and $\sigma^z_i \to s_0$. The inverse temperature $\beta$ is taken from the schedule in Fig. \ref{fig:sch}. Here, we joined the spin flips for particular configurations of the neighbors that happen to have equal rates into a single $L_\mu$ operator so that it would preserve the coherence between different states of the neighbors in that case. $P_{r_{\pm}}$ are the corresponding joined projectors that may have a rank higher than $1$. $r_+$ and $r_-$ iterate through all possible values of the rates $r_{\pm,s}$. We note that $r_\pm=1$ is the most frequent as well as the highest rate, with about half of all configurations. The number of terms $L_{i,1,+}$ and $L_{i,1,-}$ is the range of index $i$, which is the total number of spins. The number of distinct Lindblad terms on each spin can be estimated as follows. From Fig. \ref{fig:lay} we see that there are 2 neighbors linked by $J_1$ and 12 neighbors linked by $J_2 =J_3$. There are $m+1$ possible values of the total spin for $m$ spins. Thus, the number of possible rates can be estimated as $3\cdot13$ (we took into account that half of them lower the energy and thus map to $1$).

Due to a very small value of $h_{TF}$, this choice of $L_\mu$ may be sufficient to reproduce the nonequilibrium states observed in the experiment \cite{nekrashevich2022reaching}, and either positive or negative result will shed light on the underlying physics of the phenomenon. A more advanced approach that is more likely to reproduce the experiment is using $L_\mu^{\text{therm}}$ as defined in \cite{chen_quantum_2023}.

The smallest size expected to show relevant physics is 25 layers of 9$\times$9 lattices (2,025 spins). As the three plateaus that are seen in the nonequilibrium magnetization are steps of about 10\% of the full magnetization range, we set the required error in the measurement of total magnetization to 10\%.

For the simulation time, the week of the actual experiment can be accelerated by increasing the bath coupling. Let us assume that a 100-fold increase is possible. We will further simplify the task, requiring that the simulation needs to model just $T=1$s duration of the experiment. The precise sweep of the magnetic field is not necessary, so we just need to increase it in steps for 10 steps to observe the same behavior. The precision of each step is 0.01.

\subsection{Metal-Insulator transition material search application problem}

We will now specify an application instance corresponding to the entire material search in the MIT space. We estimate that the search will need to cover 1000 models of materials at 1000 parameter points each. Indeed, currently, $\sim 100$ materials with MIT are known \cite{georgescu2021database}, and we take one order of magnitude higher number as the size of the search space. One thousand parameter points will contain $10$ values of temperature, doping, and orbitals kept in the downfolding of the ab initio model.

We suggest starting the search with Hubbard model transport on a 10$\times$10 grid, which we will use in the Landauer setup \cite{landauer1987electrical} (attaching fermion sources and sinks on the left and right edge) as the foundation for the resource estimate. 4$\times$4 grid has been done classically via exact diagonalization and Kubo formula \cite{zemljivc2005resistivity}, while 10x10 is large enough to be classically challenging and make the finite size effects negligible. A simulation of transport in an integrable model shows the non-universal effect of sources and sinks to extend only 2-3 sites in \cite[Fig. 5]{prosen2008third}.

We note that the Kubo formula computation \cite{zemljivc2005resistivity} can be translated into a closed system quantum algorithm that measures the Fourier transform of the correlation function and subsequently integrates it. This is in contrast to the open system Landauer approach we pursue. It may be possible to obtain a comparable performance of this algorithm by performing all the integration inside the quantum algorithm, but naively the open system algorithm requires doing the dynamics just once to obtain the steady state and measure the flow directly, while the closed system algorithm requires performing the dynamics multiple times and integrating the results. Also, the closed system algorithm will need to start with preparing the thermal state, which may have a comparable cost to our open system evolution. We will only do the resource estimate for the open system case.

We expect that the material search will start with simple problems like the $10\times10$ example and eventually go into the ab initio investigation as guided by the experts analyzing the results of the previous runs. Such an investigation may require up to $300$ orbitals per site. Indeed, $300$ orbitals per site were used in \cite{cui2023ab} for ab initio prediction of superconducting order (which then was used to effectively predict $T_c$). Besides the Hubbard model, we will propose a simple model that can be used as a way to estimate the resources required for the average case of the material search. This model will contain $\sim 150$ orbitals per site as the middle ground between the low-end and high-end costs. 

Our Hubbard application instance is a direct generalization of the problem considered in \cite{zemljivc2005resistivity}. While a $t$-$J$ model is considered in that work, it is known to be a large-$U$ approximation of the Hubbard model, which is easier to simulate on a quantum computer. Fermi-Hubbard Hamiltonian describes spin-$1/2$ fermions on a square lattice:

\begin{equation}
    H = -\sum_{i,j,s} t_{ij} {\tilde{c}}_{js}^\dag {\tilde{c}}_{is} + U \sum_{i}n_{i,\uparrow}n_{i,\downarrow} \label{eq:HM}
\end{equation}

Each lattice site contains two fermionic modes labeled by $s\in\{\uparrow,\downarrow\}$. The hoppings contain nearest-neighbor and next-nearest neighbor terms $t$ and $t' =-0.15 t$. The repulsion $U = \frac{20 t}{3}$ and \cite{zemljivc2005resistivity} take $t =0.3 eV$. These choices are typical for downfolded models of cuprates. We will aim for a direct comparison with \cite{zemljivc2005resistivity}. Though they use the periodic boundary conditions, we expect that the finite-size effects will not be as prominent at larger system sizes; therefore, we will use open boundary conditions. The authors compute conductivity via the Kubo formula as a function of frequency $\omega$ and inverse temperature $\beta$, averaged over various boundary conditions, and apply broadening by $\epsilon =0.07t$. This broadens the delta functions present due to coherent no-scattering transport. After that broadening, they evaluate the resistivity as $6.6 \text{A} / (\text{conductivity}(\omega=0)\cdot e_0^2/\hbar)$ and find good agreement with experimental values.

The initial state of the model above is characterized by a parameter $c_h$ describing the number of holes in the system, which is a conserved quantity of the Hamiltonian. For example, $c_h = 3/20$ means that the system of 20 sites has enough electrons to fill every site with one electron of some spin value, except for three unoccupied positions. As the initial states are not required to be in equilibrium, any fermionic state with this fraction of excitations can serve as an initial state. For example, one may start with each position single-occupied, half in $\downarrow$ and half in $\uparrow$ spin state, and then destroy electrons on the first $c_hL^2$ sites for an $L\times L$ lattice. To formulate an open system problem equivalent to the Kubo formula used in \cite{zemljivc2005resistivity}, we set up sources and sinks for the electrons on two opposite sides of the square lattice, thus breaking the particle-hole symmetry. We note that at half-filling a product of particle-hole and reflection remains a symmetry of the open system dynamics. To derive the thermal Lindblad operators for sources and sinks for general $c_h$ and $\beta$, one needs to specify the spectral density $\gamma_{L,R}(\omega)$ of the creation-annihilation operators in our system as a function of temperature and chemical potential. This may be done either as a separate task on a quantum computer or classically using a smaller size patch or a free fermion approximation of the system to compute correlation functions and spectral density. Either way, an efficiently computable interpolation for the spectral density needs to be used. One then would use the obtained spectral density to build the Lindblad operators $L_\mu^{\text{therm}}$ on the left and right boundary from the hopping terms split by the system boundary, with spectral densities for two different values of chemical potential on the left and the right side, according to the general prescription of \cite{chen_quantum_2023}.

For our resource estimate, we will use the simplest possible case $L_{\text{left}}=c$ and $L_{\text{right}}=c^\dag$ on every site and spin type of the left and right edges of the square lattice, and compare it to the high temperature of the results \cite{zemljivc2005resistivity}. If this system equilibrates for a time $T$, the transport obtained would be roughly equivalent to the broadening of the delta functions in the closed system Kubo formalism used in \cite{zemljivc2005resistivity} by $1/T$. This gives $T = 1/\epsilon = 1/0.07t$, which we further adjust by a factor $10/4$ of linear increase in the system size:
\begin{equation}
    T_{\text{min}}\approx 36
\end{equation}
The full equilibration time may be much longer. The only handle we have on it is the lower bound from the fact that the information propagating linearly should reach the other end of the system, which motivates linear rescaling. 

The local temperature and chemical potential can be measured by fitting an observable, such as the local part of $H$, or the density $n$, to its value for a small system size in a Gibbs state obtained via exact diagonalization. The electric field $E$ can be found as the gradient of the chemical potential. Finally, one measures the current operator $j$ (one boundary part of $[H,n]$ where $n$ is the density in some region of the system). According to Maxwell's equations, the conductivity $= j/E$. In other words, the task for the quantum simulation is to measure n and a group of local terms of H in some portion of the system enough to estimate the gradient of the underlying chemical potential, as well as the current $j$. For simplicity, we will assume that all of these measurements can be performed in a single shot since all of these operators are local, and the system is much larger than the 4$\times$4 size accessible classically. The non-commuting terms in the Hamiltonian can be measured at various locations perpendicular to the current direction. After that, one point can be placed on the plot of resistivity = $1/$conductivity. To compare with the 4x4 results of \cite{zemljivc2005resistivity}, one may need to convert it to physical units using the lattice spacing of $1.65\text{A} $ and the quantum of conductance.

 {\emph{Classical state-of-the-art}} Recent ab initio numerical investigation by \cite{cui2023ab} shows that even the high-$T_c$ transition temperature can now be predicted by numerical techniques. Since the metal-insulator transition in application-relevant materials happens at higher temperatures, there is a possibility that classical techniques may predict it fully ab initio as well. The apparent absence of publications on this topic may be due to conceptual difficulties with computing a transport coefficient numerically from first principles (unknown disorder, different assumptions of Kubo and Landauer approaches, apparent divergence of finite size DC conductivity without disorder), not due to computational hardness. We also note that machine learning may be successful in the material search for MIT materials, with some initial results presented in \cite{georgescu2021database}. Finally, the cost of searching for MIT transitions experimentally at room temperatures may not be very high for cheap fabrication methods. What makes large throughput experimental material searches unfeasible in condensed matter is that the material properties are highly sensitive to fabrication quality, and such a search would also require automated optimization of fabrication technology for each material. Another concern is that many of the candidate materials are toxic. 

 {\emph{Average application instance for material search}} While simulating Eq. (\ref{eq:HM}) will stimulate theory research, it is not directly useful for an ab initio material search. Ab initio prediction \cite{cui2023ab} of superconductivity required keeping 150 orbitals per site without downfolding. For the back-of-the-envelope estimate of the cost of our Landauer approach when used for such ab initio predictions, we assume that a typical material search will involve simulating instances on a 10$\times$10 grid where each site is also described by $\sim 10^2$ orbitals. Among ab initio systems of that size, the one with the most optimized quantum simulation algorithms is FeMoCo (one of the latest improvements is in \cite{lee2021even}). We will construct the following artificial Hamiltonian that is meant to have the same features as an average instance of the material search: consider a 10$\times$10 grid of noninteracting FeMoCo Hamiltonians. On each site of the boundaries, Lindblad operators $L_\mu^{\text{therm}}$ are applied that prepare a low-temperature state corresponding to an excess or absence of some of the electrons. We have freedom in their choice, and one way to guarantee an approach to appropriate Gibbs distribution is to follow \cite{chen_quantum_2023}. The appropriate norm of the FeMoCo Hamiltonian (not $\|H\|$, but we will use them interchangeably for the back-of-the-envelope estimate) is $\lambda \sim 1500$ \cite{lee2021even}, and the cost of its block-encoding is $\sim 17,000$ \cite{harrigan2024expressing}. We rescale the required evolution time by $150$ orbitals:
 \begin{equation}
    T_{\text{Av.}}\sim 5000
\end{equation}
The estimates for this instance will quickly result in very high resource costs, so we leave it for future work to develop new simpler forms of $L_\mu^{\text{therm}}$ that have low-temperature Gibbs state as the steady state and algorithms compatible with this model.

 \section{Quantum Implementation}
\label{sec:qi}
For both applications we consider, we require an algorithm that would simulate an evolution (for time $T$, to precision $\epsilon$) of state $\rho$ under an open system dynamics given by a Hamiltonian $H$ and Lindblad generators $L_i$. In both cases, $H$ is a sum of local terms, and $L_i$ are local. Many algorithms for this task have been proposed. We will briefly mention other approaches before describing the approach we believe to be optimal:
 \begin{itemize}
     \item Kraus operators corresponding to the Lindblad evolution for a short time can be used together with various quantum algorithms to implement quantum channels given Kraus operators. For example, \cite{hu2022general} were able to realize a circuit compact enough for NISQ devices. That approach appears to have highly unfavorable scaling with the evolution time.
     \item Any approximate implementation of $e^{\delta\mathcal{L}_i}$ for a small $\delta$ and the Lindbladian with one generator $L_i$ can be used in a Trotter formula, which we do in App. \ref{app:ex} for the example of an exact resource count calculation. The scaling with time is $\sim T^2$. The power is improved in \cite{ding2024single}, approaching $1$ as the order of the Trotter formula is increased.
 \end{itemize}
What we are interested in is commonly referred to as quantum signal processing (QSP) scaling, which is the $T$log$(1/\epsilon)$ scaling of the algorithm runtime with evolution time $T$ and precision $\epsilon$, possibly at the cost of extra powers of the system size in front.

{\bf{Lindbladian QSP:}}
QSP scaling for open system dynamics was originally achieved in \cite{cleve2016efficient}. We will follow the approach of \cite{chen_quantum_2023} that allows further optimizations during the block encoding of $\mathcal{L}$.

\begin{figure}[t]
\centering
\includegraphics[width=0.75\columnwidth]{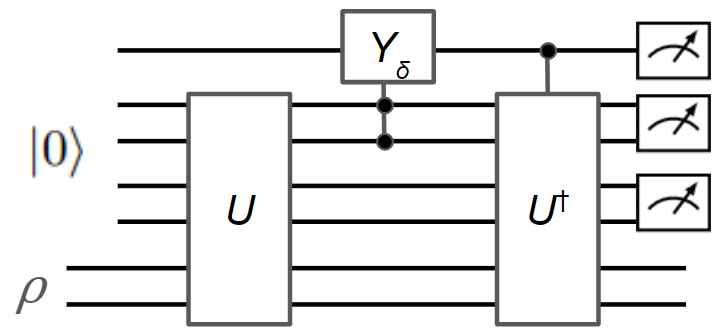}
\caption{A simple circuit approximately implementing an open system evolution $e^{\delta \mathcal{L}}\rho$ via the block-encoding of a Lindbladian $U$. The gate $Y_\delta= e^{-iY\text{arcsin}\sqrt{\delta}}$}
\label{fig:evo}
\end{figure}

First, consider a simple (not achieving QSP scaling) evolution step in the language of \cite{chen_quantum_2023} via the circuit in Fig. \ref{fig:evo}.

Here $U$ is a block encoding of a Lindbladian containing all the Lindblad operators conditioned on the value of the ancilla. Iterating this is the simplest way to simulate the open system dynamics. A more advanced algorithm also found in \cite{chen_quantum_2023} achieves QSP scaling for the number of gates. Specifically, Theorem III.2 of \cite{chen_quantum_2023} uses a circuit more advanced than shown above, a compression algorithm on the number of applications of the Lindbladian in the weak measurement version of QSP. To our knowledge, there are no QSP scaling algorithms for time-dependent Lindbladians. Developing them is an important direction for future work. A QSP scaling algorithm for a closed system time-dependent Hamiltonian on a lattice was demonstrated in \cite{haah2021quantum}.

While Lindbladian QSP may have an optimal scaling (except for the factor of the system size due to block encoding), for small errors and evolution times other algorithms, such as Trotter, may have an advantage in gate count. As the gate count for both is only available in terms of big-O notation, these algorithms require new exact resource estimates. Our main results rely on order-of-magnitude estimates that do not keep track of the exact prefactor. We also present preliminary, heavily suboptimal exact resource estimates for Trotter in App. \ref{app:ex}.

\section{Resource estimation for proposed quantum algorithms}
\label{sec:re}

Below, we provide the order of magnitude resource estimates.

\subsection{Abstract Logical}
\subsubsection{Lindbladian QSP}
Theorem III.2 of \cite{chen_quantum_2023} lists the following requirements for simulating the open system time evolution for time $t$ to precision $\epsilon$. It relies on the block encoding of the Lindbladian $\mathcal{L}$ governing the evolution, i.e., a unitary $U$ such that all $M$ Lindblad generators $L_\mu$ can be found as blocks in the upper left corner of it, and the block encoding $V$ of the system Hamiltonian: 

\begin{equation}
    U = \left(
\begin{array}{c|c}
L_1  & \cdot\\
 L_2 &\cdot\\
  \dots &\cdot \\
   L_{M} & \cdot\\
\hline
\cdot& \cdot \\
\end{array}
\right),  \quad   V = \left(
\begin{array}{c|c}
H & \cdot\\
\hline
\cdot& \cdot \\
\end{array}
\right)\label{upBlock}
\end{equation}
Let $c$ be the number of ancillas used by $U$ and $V$. The quantum algorithm uses $O((c + \text{log}((t + 1)/\epsilon)) \text{log}((t +1)/\epsilon))$ resettable ancillas, $O((t + 1) \text{log}((t + 1)/\epsilon) /\text{loglog}((t + 1)/\epsilon))$ controlled uses of $U$ and its conjugate, same amount of uses of each of $V$ that block encodes the Hamiltonian,  and $O((t + 1)(c + 1)\text{polylog}((t + 1)/\epsilon))$ other 2-qubit gates. It remains unspecified how many of those gates are Clifford, as the details of the composition of the circuit outside $U$ are not fully presented in \cite{chen_quantum_2023}, instead appearing in various references therein. We note that there can be no more than $O((t + 1)(c + 1)\text{polylog}((t + 1)/\epsilon))$ non-Clifford gates and arbitrary rotations outside $U$. As we will be able to perform a detailed estimate of the cost of (controlled) $U$, it will become apparent that the non-Clifford count $O((t + 1)(c + 1)\text{polylog}((t + 1)/\epsilon))$ is subleading. Our strategy will be to keep only the leading terms in the cost and replace big-O notation with a factor 1. This will be sufficient for the initial analysis we perform. As an example of rigorous resource estimation, we analyze a suboptimal evolution step in Fig. \ref{fig:evo} in a Trotter evolution algorithm in App. \ref{app:ex}.

\subsubsection{Block encoding for a Lindbladian}

\begin{figure}[t]
\centering
\includegraphics[width=\columnwidth]{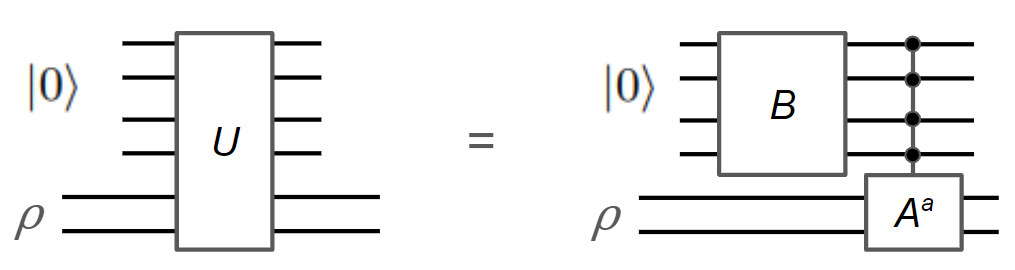}
\caption{A circuit for the block-encoding of a Lindbladian $U$ given by its unitary Lindblad generators.}
\label{fig:GENcir}
\end{figure}

The general way to construct $U$ defined in Eq. (\ref{upBlock}) follows the structure of what is denoted $V_{jump}$ in \cite{chen_quantum_2023}. A set of unitary operators $A^a$ is block encoded in it as follows:
\begin{equation}
    V_{jump} = \left(\sum_{a\in A} |a\rangle \langle a| \otimes A^a \right)\cdot (B\otimes {\mathds{1}}_{\text{sys}}),
\end{equation}
where the operator $B$ prepares a uniform superposition $B| 0\rangle = \sum_{a\in A} \frac{1}{\sqrt{|A|}}|a\rangle $  following \cite{shukla2024efficient}.  For the simplest case where $L_a =A^a$, which are unitary, the block encoding of the Lindbladian is just $U = V_{jump}$ illustrated in Fig. \ref{fig:GENcir}.
Since in our case the operators $L_\mu$ are not unitary and have $O(1)$ cases of different coefficients in front, we will need to utilize an $O(1)$ number of extra ancillae for the block encoding:
\begin{equation}
    U= \left(\sum_{\mu=1}^M|\mu\rangle \langle \mu| \otimes U_\mu\right)\cdot (B\otimes {\mathds{1}}_{\text{anc}}\otimes {\mathds{1}}_{\text{sys}})
\end{equation}
where $U_\mu$ is a block encoding of an individual Lindblad operator $L_\mu$:
\begin{equation}
    U_\mu =  \left(
\begin{array}{c|c}
L_\mu & \cdot\\
\hline
\cdot& \cdot \\
\end{array}
\right).
\end{equation}
Note that the expression $\sum_{\mu=1}^M|\mu\rangle \langle \mu| \otimes U_\mu$ is used as a shorthand. First of all, if the register $|\mu\rangle$ is represented in qubits and there is a mismatch between the dimension of the register and the total number of Lindblad terms, we use  $\sum_{\mu=1}^M|\mu\rangle \langle \mu| \otimes U_\mu \oplus{\mathds{1}}$, but we will omit $\oplus{\mathds{1}}$ for brevity. Second, the specific circuit implementation of $\sum_{\mu=1}^M|\mu\rangle \langle \mu| \otimes U_\mu$ may not utilize controlled $U_\mu$. Indeed, below we will utilize symmetries between different $U_\mu$ to substantially improve the depth of $\sum_{\mu=1}^M|\mu\rangle \langle \mu| \otimes U_\mu$ at the cost of $\sim$system size number of ancillae.

The cost that will dominate is the leading term of system-size dependence coming from the operator $\sum_{\mu=1}^M|\mu\rangle \langle \mu| \otimes U_\mu$, where $U^a$ is the block-encoding of the Lindblad generator $L_a$. Let us briefly describe the other operator $B$ and the related rescaling of the Lindblad generators. Depending on the norms of the operators $\|H\|, \|L_\mu\|$, rescaled version $ H/\alpha_H, L_\mu/ \alpha_L$ appear in the block encoding Eq. (\ref{upBlock}). The time has to be rescaled as $t \to \alpha t, ~ \alpha = $max$(\alpha_H, \alpha_L)$. We will now compute the rescaling factors $\alpha_H, \alpha_L$. The Hamiltonian one is determined from unitarity of $V$, thus $|\langle 0 | V|0\rangle| \leq 1$ and we can take $ \|H\|/\alpha_H = 1$. Another comes from the specific construction above. Let $U_\mu$ encode $L_\mu/\|L_\mu\|$. The operator $B$ contains a matrix element $\sum_\mu c_\mu |\mu\rangle \langle 0|$ such that $\sum_m |c_\mu|^2 = 1$. The $U|0\rangle = \sum_\mu c_\mu /\|L_\mu\| |\mu\rangle  \otimes U_\mu$ has the following block structure:  
\begin{equation}
    U = \left(
\begin{array}{c|c}
\frac{c_1}{\text{max}(1, \|L_1\|) }  L_1  & \cdot\\
\frac{c_2}{\text{max}(1, \|L_2\|) } L_2 &\cdot\\
  \dots &\cdot \\
  \frac{c_M}{\text{max}(1, \|L_M\|) }  L_{M} & \cdot\\
\hline
\cdot& \cdot \\
\end{array}
\right)
\end{equation}
The solution is to pick $c_\mu = \frac{1}{\alpha_L} \|L_\mu\|$, which leads to
\begin{equation}
   \alpha_L^{-2} \sum_\mu \|L_\mu\|^2 =1
\end{equation}
The time rescaling factor is then:
\begin{equation}
    t \to \alpha t, \quad \alpha = \text{max}(\|H\|, \sqrt{\sum_\mu \|L_\mu\|^2})
\end{equation}
The Hamiltonian contribution is larger for both of our application instances. Ca$_3$Co$_2$O$_6$, $\|H\| \sim 3.2\cdot 2025$, $\alpha \approx 6500 $ . For Hubbard,  $\|H\| \sim 11 \cdot 100$, $\alpha \approx 1100$.

\subsection{Compiled Logical}

\subsubsection{Single Lindblad generators}
\label{sec:1l}

For Ca$_3$Co$_2$O$_6$, the Lindblad terms (Eq. (\ref{CaLi})) contain $\sigma^\pm$ multiplied by projectors onto the state of a constant number of neighboring spins. For the Hubbard problem, after the Bravyi-Kitaev \cite{bravyi2002fermionic} or Jordan-Wigner transformation from fermions to qubits, the creation-annihilation operators would have a logarithmic weight, containing a product of Pauli's and one $\sigma^\pm$. We will now briefly outline a possible (not necessarily optimal) block-encoding of those operators. First, we note that $\sigma^\pm$ can be block encoded into a 2-qubit (Clifford) permutation:
\begin{align}
    U_+ = X_0 \, \text{SWAP} = \left(
    \begin{array}{cc|cc}
        0 & 1 & 0 & 0 \\
        0 & 0 & 0 & 1 \\
        \hline
        1 & 0 & 0 & 0 \\
        0 & 0 & 1 & 0 
    \end{array}
    \right) =\left(
\begin{array}{c|c}
\sigma^+ & \cdot\\
\hline
\cdot& \cdot \\
\end{array}
\right), \\ U_-= X_1 \, \text{SWAP} = \left(
    \begin{array}{cc|cc}
        0 & 0 & 1 & 0 \\
        1 & 0 & 0 & 0 \\
        \hline
        0 & 0 & 0 & 1 \\
        0 & 1 & 0 & 0 
    \end{array}
    \right) = \left(
\begin{array}{c|c}
\sigma^- & \cdot\\
\hline
\cdot& \cdot \\
\end{array}
\right).
\end{align}
The encoding is in the top left block. Note here we take the convention $\ket{\uparrow} = \ket{0}, \ket{\downarrow} = \ket{1}$, and $\sigma_- \ket{\uparrow} = \ket{\downarrow}$ etc.

The evolution algorithms \cite{chen_quantum_2023} also also require a controlled version $CU_\pm$. As counting required T-gates and depth is fairly technical and can distract the reader from our main message, we discuss all the relevant methods for that in App. \ref{app:howtocount}, and only quote the results here. $CU_\pm$ in particular require 4 T-gates and depth 14.

The block encoding of creation-annihilation operators contains the Pauli's on system qubits multiplying $U_\pm$, which results in the same non-Clifford cost $0$ and $4$ for $U_\mu$ and $CU_\mu$ respectively. The depth is $4$ for $U_\mu$ as all the Pauli's on $\sim100$ (on average) other qubits can be applied in parallel. For $CU_\mu$, one first needs to prepare up to $199$ long register with the repeated value of the control, do every $CZ$ in parallel, and then undo the register. The average depth of that is $\text{log}199 \sim 8$ and can done in parallel with the CSWAP. We will later need to control $CU_\pm$ on the type information $\pm$. In App. \ref{app:howtocount} we show that $C^2U_\pm$ can be implemented with $12$ T-gates and average depth $37$.

For lower temperatures, we expect to numerically find low-cost block-encodings of Lindblad generators among a family where a local raising or lowering operator is multiplied by functions of the neighboring occupation numbers and terms like $c^\dag_i c_j$. These terms will only require diagonalization of the system in a local neighborhood. The more complex unit cell of the average case MIT material search requires even more optimization, as there is no notion of locality between orbitals. The straightforward application of the thermal Lindblad generators suggested in \cite{chen_quantum_2023} results in unrealistic (while still polynomial) resource counts, as we show in App. \ref{app:av1019}.

In the Lindblad generator for  Ca$_3$Co$_2$O$_6$, instead of Pauli's, projectors $P_{r\pm}$ on various states of the neighbors multiply $\sigma^\pm$. They also need to be block encoded using additional ancillae. Note that since $P_{r\pm}$ is a projection on $14$ spins, one can use one additional ancilla to form a unitary:
\begin{equation}
    P_{r\pm} \otimes {\mathds{1}}  + (  {\mathds{1}}-P_{r\pm})\otimes X
\end{equation}
A rank-1 projector, for example, is block-encoded in the following matrix:
\begin{equation}
\left(
\begin{array}{ccccc|ccccc}
1 & 0 & \dots& 0& 0 &0 &0 & \dots&0 &0\\
0 & 0& \dots& 0 & 0 & 0 & 1& \dots &0& 0 \\
0 & 0& \dots& 0 & 0 & 0 & 0& \dots&0 & 0 \\
\dots & \dots & \dots& \dots & \dots & \dots &  \dots &\dots & \dots & \dots\\
0 & 0& \dots & 0& 0 & 0 & 0 &\dots &0 & 1 \\
\hline
0 & 0& \dots & 0& 0 & 1 & 0 &\dots &0 & 0 \\
\dots & \dots & \dots& \dots & \dots & \dots &  \dots &\dots & \dots& \dots \\
0 & 0& \dots & 1& 0 & 0 & 0 &\dots &0 & 0 \\
0 & 0& \dots & 0& 1 & 0 & 0 &\dots &0 & 0 \\
\end{array}
\right)
\end{equation}
There are multiple ways to implement this unitary, which is essentially a $\tilde{C}X$ conditioned on the states of the qubits to be in the support of $P_{r\pm}$. For a projector rank $k$ with no additional structure, this unitary can be implemented using $k$ applications of $C^{14}X$ (or $2^{14}-k$ applications, whichever is smaller). $k$ varies between $L_i$, so we take an average value $2^{14}/39 \approx 420$. In other words, it would require $420$ applications of $C^{14}X$. The controlled application would utilize $C^{15}X$ instead. As these numbers are very large, we should instead utilize the structure of $P_{r\pm}$: the spin configurations correspond to particular values of the sum $|J_1|\sum_{i\in \text{red}}s_i + 0.1|J_1|\sum_{i\in \text{green,~blue}}s_i$ of $14$ binary variables, $2$ of which are multiplied by 10. As we didn't count for possible cancellations, we should simulate the binary adder circuit on groups of $12$ and $2$ and condition the application of $X$ on the result of those additions. The adder is a circuit that prepares a sum of 2 numbers stored as $m$ qubits each in binary. Following \cite{gidney2018halving}, when this circuit is used for an operation conditioned on the bits of the sum and then undone, the total cost is $\sim 4m$ T-gates, and the total depth is $\sim 18m$. In our case, instead of adding two numbers, we add single-bit numbers $n=12$ and $n=2$ times, so the largest $m$ needed is $m=3$ for the sum of parts of $12$ spins. No matter how we group them, the total $\sum m =20$, which results in the counts:
\begin{equation}
    \text{T count} \approx 80, \text{depth} \approx 360
\end{equation}
As the results of the addition can be stored in 6 qubits, The block encoding of $P_{r\pm}$ also includes $C^{6}X$, and the block encoding of $P_{r\pm}\sigma ^{\pm}$ includes additional Cliffords of $\sigma ^{\pm}$ that are done in parallel.  For $CU$, we will duplicate the ancilla with a CNOT (depth $+2$ for it and its undoing) and apply $C^{7}X$ and CSWAP from $\sigma^-$ in parallel. We can translate these gates into T-counts and depths ($C^{7}X \rightarrow 24$ T-gates, depth 57, $C^{6}X \rightarrow 20$ T-gates, depth 53) following App. \ref{app:howtocount}. The total counts are:
\begin{itemize}
    \item $U_\mu$: T-count  $100$, depth $413$
    \item $CU_\mu$: T count $108$, depth $419$
\end{itemize}
More generally, there are $39$ types of generators on each site, and let the number of sites be $M$. If we need to control on $k=$log$_2 39M$ other qubits, the counts for $C^kCU_\mu$:
\begin{itemize}
    \item  T count: $80$+[C$^{7+k}$X+ C$^{2+k}$X count] $= 108+8k$ \item depth: $362$+[C$^{7+k}$X depth]$=393+32\text{log}_2(k+1)$
\end{itemize}

We collect the results as follows:
\begin{center}
\begin{tabular}{ |c|c|c| } 
 \hline
T-count; depth & Hubbard & Ca$_3$Co$_2$O$_6$ \\ 
\hline
 $U_\mu$ & $0$; $4$ & $100$; $418$ \\ 
$CU_\mu$ & $4$; $14$ & $108$; $427$ \\ 
$C^{\text{type}}CU_\mu$ & $12$; $37$ & $108+8k$; $393+32\text{log}_2(k+1)$ \\ 
 \hline
\end{tabular}
\end{center}
For this table, the type is encoded in the $k=1$ qubit register for Hubbard.

Note that only an $O(1)$ number of ancillae is required for the implementation of the block encoding of the individual Lindblad operators, so it will be neglected in our estimate.

\subsubsection{Block-encoding of full Lindbladian}
As we have seen in the previous section, $U$ involves the select step $\sum_{\mu=1}^M|\mu\rangle \langle \mu| \otimes U_\mu$, which will turn out to be the leading contribution to the cost, even in the simplest possible case when $U_\mu$ is a single-qubit Pauli $A_\mu$, which we will use for scaling estimates here before applying the intuition we gain to more complicated $U_\mu$ of the application instances. The operation $\sum_{\mu=1}^M|\mu\rangle \langle \mu| \otimes A_\mu$ can be compiled in different ways. Naively, one can use $M$ consecutive log$_2M$-controlled application of $A_\mu$. The T-gate count of an individual such gate would be $\sim4$log$_2M$, ~ CNOT-count $\sim 4$log$_2$log$_2M$ and depth $\sim 16$log$_2$log$_2M$\cite{he2017decompositions}, which results in total T-count and depth of $U$ with the leading terms:
\begin{align}
    \text{T-count }\sim4M\text{log}_2M, \quad \text{CNOT-count }\sim 4M\text{log}_2M \label{Ustan} \\
    \text{depth }\sim 16M\text{log}_2\text{log}_2M
\end{align}
Now consider $CU$ that is applied in the time-evolution algorithm. If the system size $n$ is comparable to $M$, it is optimal to add control to every gate in $U$: CNOT-count maps to CCZ-count and T to CT. If in $CU$ several gates controlled on the first qubit need to be applied in parallel, we use a tree of CNOTs to obtain enough classical repetitions of it at the start of $CU$, use them to control the gates in parallel, and undo the CNOTs. This step will only add an $M$-independent constant to depth, which is subleading and will be dropped. The counts for $CU$ are as follows:

\begin{align}
    \text{T-count }\sim36M\text{log}_2M,  \\
    \text{depth }\sim cM\text{log}_2\text{log}_2M \label{trUstan}
\end{align}
Here $16\leq c\leq 208$ depending on the positions of CNOTs and Ts.

We will now discuss the possibilities for depth reduction. For instance, at the expense of increasing the number of ancillae to $M$, it would be possible to instead use one-hot encoding for ancillae and single controlled gates for $|\mu\rangle \langle \mu| \otimes U_\mu$ in constant depth, as many of the operators $U_\mu$ act on independent sets of qubits and may be parallelized. The {\emph{Prepare}} step $B$ for such a one-hot encoding, however, would still require $\sim M$ depth and T-cost. These ideas are further refined in \cite{babbush2018encoding}, yet we do not apply those results here. Instead, we utilize the translation invariance of $U_\mu$ if such is present. Indeed, for our scaling estimate with $U_\mu= A_\mu$, assume that $A_\mu$ are translations of the same single-qubit Pauli $A$, and $\mu$ indexes the lattice site.  If we change our point of view, exactly one generator is always applied by $\sum_{\mu=1}^M|\mu\rangle \langle \mu| \otimes A_\mu$, what changes is the location. As the location is encoded in the ancillae $|\mu\rangle$,  we can implement a conditional SWAP circuit that brings the target location $\mu$ to a specific site, apply $A$ on it, and reverse the SWAPs. More generally, if $U_\mu$ is acting on a set of qubits in a translationally invariant way, one swaps the whole set. As this is our original contribution, we describe the construction for a single qubit operator $A_\mu$ in detail below and also discuss the translational invariance in the case of fermionic $c,c^\dag$. The proposed circuit for a system size $M$ has $O(M)$ T-count and $O($log$M)$ depth (and T-depth), which is a substantial improvement in the parallelism of T-gate applications compared to the approach above.

We will note that in the case of our applications $L_\mu$ is more complex than a single-qubit Pauli for Ca$_3$Co$_2$O$_6$, and for Hubbard $M=20$ is just not large enough, so neither of those problems benefit from the depth reduction. Both, however, strongly benefit from the reduction in the number of circuits for $U_\mu$ in the final algorithm that is enabled by this translation invariance trick.

\subsection{Translation-invariant block-encoding}
First, consider the ancilla $|\mu\rangle$ storing the position information (for now, there is only one type, so the entire $|\mu\rangle$ is used for position). For $M$ different positions, $|\mu\rangle$ encodes the position in binary using $\text{log}M$ qubits. For the first step, we need to map $|\mu\rangle$ into a state on $O(M)$ qubits as follows. For the qubit storing $k$'th bit of position information, we create $M_k$ classical repetitions of that bit via a tree of CNOTs. Such a tree has $\sim M_k$ CNOTs and depth $\text{log}M_k$. The first bit, storing the information about which half of the system the operator is applied to, is repeated $M/2$ times, the next ~-- $M/4$, and so on, for a total CNOT count $\sim M$ and depth $\sim\text{log}M$ used to generate 
the classical repetitions for each qubit in $|\mu\rangle$ (on a qubit in a $|+\rangle$ state, this circuit is familiar as the cat state preparation). 

\begin{figure}
    \centering
    \includegraphics[width=\columnwidth]{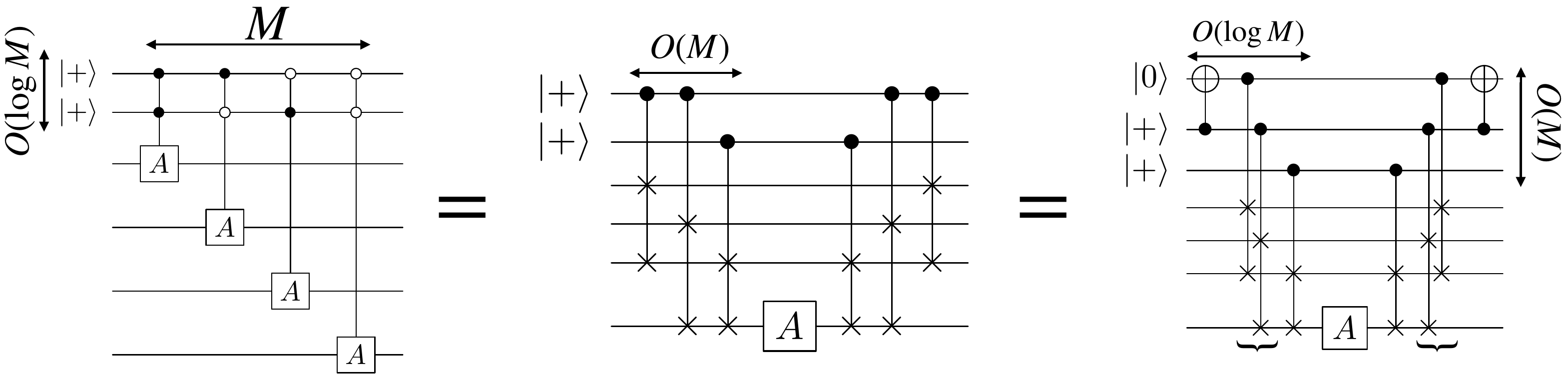}
    \caption{Three equivalent ways of preparing $\sum_{\mu=1}^M \ket{\mu}\otimes A_\mu \ket{\psi}$, where $A_\mu$ is the single qubit unitary $A$ applied on qubit $\mu$. In this example $M=4$. The left/middle figures are implemented with depth $O(M)$ and ancilla cost $O(\log M)$, whereas the rightmost has an ancilla cost $O(M)$ and a depth $O(\log M)$. The curly braces in this figure indicate the CSWAPs can be implemented in parallel. The CNOT comes from the GHZ preparation circuit (see main text).}
    \label{fig:block3}
\end{figure}

We then initiate the SWAPs with the goal to swap the qubit in position $\mu$ to the fixed position $\mu=1$, apply a gate $U_\mu$, now not controlled by anything, and swap back.  
We see that for single-qubit $L_\mu$ that are all the same, implementing $\sum_{\mu=1}^M|\mu\rangle \langle \mu| \otimes U_\mu$ would only require $\sim M$ SWAPs conditioned on one qubit each, and a single deterministic application of $U_0$. This is illustrated for $U_a = A_a$ in Fig.~\ref{fig:block3}.

The leftmost figure of Fig.~\ref{fig:block3} is the usual way to implement block encoding, which is the typical choice when the unitaries have no structure. In the case of translation invariance, we can, however, rewrite the circuit using only CSWAPs and a single use of $A$ (not controlled). 
This is essentially an application of binary search; if the ancilla state is $\ket{\mu}$, to apply $A$ on qubit $\mu$, one can use $O(\log M)$ swaps to move $A$ onto the correct wire. Ancilla bit $k$ has $O(M/2^{k})$ CSWAPs acting. Taking into account all required control swaps means this circuit has depth $O(M)$ and uses $O(\log M)$ ancillae. 

This circuit depth can be reduced to $O(\log M)$, and the expense of more ancillae ($O(M)$), which is the rightmost subfigure in \ref{fig:block3}. The observation is that the swaps in the system register, conditioned on a particular ancilla, act on disjoint pairs of qubits. It is, therefore, possible to first encode the ancilla state using more qubits, allowing application of these CSWAPs in parallel: the $k$'th bit is encoded using CNOTs as $\ket{\mu_k} \rightarrow \ket{\mu_k}^{\otimes M_k}$, where $M_k=M/2^k$ (for simplicity, assume $M$ is a power of 2). 
In Fig.~\ref{fig:block3} only one CNOT is required, but in general it can be prepared with a circuit of depth $O(\log M_k)$ \cite{Cruz_2019}. Since each of these GHZ preparation circuits (for each bit of $\mu$) can be implemented in parallel, the depth of this step is $O(\log M)$.

Thanks to our repeated position register, each of the CSWAPs corresponding to that bit of $|\mu\rangle$ can be done in parallel, so the depth in terms of CSWAPs is $\sim2\text{log}M$. For the count of $U$, we get:
\begin{align}
    \text{CSWAP-count }\sim 2M, \quad \text{CNOT-count }\sim 2M \\
    \text{depth of CNOT part}\sim 2\text{log}_2M \\
     \text{depth of CSWAP part}\sim 2\text{log}_2M
\end{align}
In the circuit for $CU$, we again apply control to every gate, assuming the system size is comparable to $M$. The two gates we used are turned into Toffoli and CCSWAP. The resulting counts for $CU$ are:

\begin{align}
    \text{T-count }\sim 10M, \quad \text{depth }\sim 58\text{log}_2M \\
    \text{ancilla count} \geq M
\end{align}
We see that while the T-count is comparable with Eq. (\ref{trUstan}), the T-depth is logarithmic instead of linear, which allows for significant parallelism. The ancilla count depends on the implementation: one can, in principle, undo the repetitions of the first bit and use the same ancilla qubits for the second. In that case, counting also the ancilla in the implementation of CCSWAP, we get ancilla count $\sim M$, thus doubling the qubit count of the algorithm. We also note that if we generalize this approach to a more complicated single-qubit $U_\mu$ compiled to multiple T gates, the $M$-independent T-count and depth of $U_\mu$ (or $C U_\mu$) applied on the qubit at $\mu=1$ is just added to the above counts, thus for sufficiently large $M$ its contribution is subleading.

We will now go over the implications of parallelism during the physical compilation. The fault-tolerant constructions used in the resource estimation \cite{beverland2022assessing} do not allow for T-gate parallelism but benefit from offloading the Clifford depth to classical computation, which also allows to ignore the connectivity constraints. An alternative construction is to keep the Clifford operations and use teleportation for all 2-qubit operations. This will allow for T-gate parallelism but multiply the depth by an additional factor of $5$ for the teleportation depth. The number of logical cycles for $CU$ is then:
\begin{equation}
    \#\text{ of logical cycles} \sim 290\text{log}_2M \text{ vs } 10M
\end{equation}
The compilation \cite{beverland2022assessing} of the same circuit consumes one T-state per cycle, thus the second estimate. For the parallelized compilation to be faster, one needs $M\geq 227$, which is one order of magnitude larger than the value in our application problems. 
The number of ancillae used by traditional block encoding scales as $c \sim$ log$_2M$, while it is $\sim M$ for the depth-optimized approach proposed here. 

We will see that Ca$_3$Co$_2$O$_6$ still benefits from the translation invariance trick due to the reduction of the number of applications of $U_\mu$, but the physical compilation does not need the T-parallelism. For the MIT instance, the special structure of the generators will allow for large parallelism, but our instance is dominated by the Hamiltonian simulation cost.

\subsubsection{Cost for Ca$_3$Co$_2$O$_6$}

In our applications, unlike the above estimate, $L_\mu$ is not a single qubit operator, and there are multiple types of $L_\mu$ on the same qubit. For Ca$_3$Co$_2$O$_6$, each spin has $14$ neighbors that need to be swapped together for this trick to work. This severely restricts the depth savings of the trick, as we will illustrate now. This 15-spin set (see Fig. \ref{fig:lay}) contains 3 red vertices, 6 green and 6 blues, so we do not expect a periodic tiling of the lattice with sets like these (the density of red vertices will be off). There is, however, a periodic tiling where 3 more red vertices are included, say 2 neighbors of the middle one and one neighbor of the bottom one. By slightly translating this tiling 18 possible ways, we get each spin to be the center. This way we can block-encode $\tilde{M}\sim 2025/18\approx 113$ possible translations of each $L_\mu$ to other tiles via just one application of the block-encoding of $L_\mu$ and the translation trick. $\tilde{M}\sim113$ is not large enough to justify compiling T gates to execute in parallel, which is why we will instead trade depth for a full translation of all spins. In this approach, the tiling will not be needed. The index $\mu$ will encode one of $2025$ possible shifts of the lattice (we will assume periodic boundary conditions for simplicity). We will now discuss the structure of the circuit $U_T$ that shifts a $9\times9\times25$ lattice by an arbitrary lattice vector. First note that a sequence of SWAPs $S_{1,2}S_{2,3}S_{3,4}\dots$ shifts the state of the qubits by 1 lattice position. $U_T$ can be in principle be implemented by using $2\cdot2025$ such sequences of length $9$, and $2025$ of length $25$, controlled on an unary encoding of $\mu$. That would cost $\sim700,000$ T-gates, which is too large. Instead note that a sequence of SWAPs $S_{1,1+k}S_{2,2+k}S_{3,3+k}\dots$ shifts the state by $k$ positions. Any shift of $2^{M}$ qubits can be encoded in an application of $M$ such sequences, with $k =1,2,4\dots 2^{M-1}$. Provided a register $\mu$ encoding the shift, the sequences can be controlled on bits of $\mu$. For our 3d lattice, we need roughly $(2\cdot2025/9)\cdot$log$9$, $(2\cdot2025/9)\cdot$log$9$ and $(2\cdot2025/25)\cdot$log$25$ sequences of length $9$,$9$ and $25$ respectively. It would cost $178,000$ T gates, and $5000$ depth. There are $39$ types of a unitary $L_\mu$ on each spin, each with costs from Sec. \ref{sec:1l}. The counts of $39 $ $C^kCU$ with $k=$log$39 \approx 5$ are:
\begin{itemize}
    \item  T count$= 39(108+8k)\sim 7,300$ 
    \item depth $=39(393+32\text{log}_2(k+1))\sim 18,500$
\end{itemize}
The total counts are:
\begin{equation}
    \text{qubits: } 2\cdot 2025; \text{ T count: 185,000; 
 ~depth 23,500. }
\end{equation}
The leading contributions to the number of qubits are the system and the ancillae controlling SWAPs.

\subsubsection{Cost for Hubbard}

For MIT instance, $c$ and $c^\dag$ each are repeated $20$ times on different sites and orbitals (however, fermionic translation invariance is not equivalent to the translation invariance of the corresponding spin operators). As the costs of $U_\mu$ and $CU_\mu$ are relatively low, and their structure is simple, we will consider both the straightforward construction of $U_\mu$ and the fermionic translation trick. The circuit for $CU_\mu$ contains two trees of CNOTs duplicating the state of the control and undoing themselves, CZ applied in parallel on the $\sim 200/2$ (on average) qubits where the chain from Jordan-Wigner lands, and two CX and CSWAP for $\sigma^\pm$. The depth is $2$log$100 +3 \approx 16$. We now intend to control it on a register $|\mu\rangle$ that has $40$ different values, represented in $\sim$log$40 \approx 5$ qubits. Only the CZ and the Toffoli inside CSWAP need to be controlled on that register, resulting in 100 C$^6$Z gates and one C$^7$Z (depth 57) gate per operator. As CZ are applied in parallel, it is possible to apply C$^6$Z in parallel as well via another duplication. Since we have already duplicated the first control, it does not require additional depth. The counts for a single C$^5U_\mu$ are then:
\begin{equation}
    \text{T-count: } 2024, \text{ depth: 72}
\end{equation}
There are 40 operators total, so the counts for $U$ are:
\begin{equation}
    \text{T-count: } 81000, \text{ depth: }2900
\end{equation}
Some savings may be obtained if we note that out of the 40 strings of Jordan-Wigner, 20 has length $<20$, and another 20 has length 180..200. Note that 180 CZ operators are applied conditioned on one bit of the $\mu$ register, thus only require C$^2$Z. The totals are 180 C$^2$Z, $40\cdot 20/2$   C$^6$Z, and $40$ C$^7$Z. The updated counts are:
\begin{equation}
    \text{T-count: } 9700, \text{ depth: }2900
\end{equation}

\begin{figure}
    \centering
    \includegraphics[width=\columnwidth]{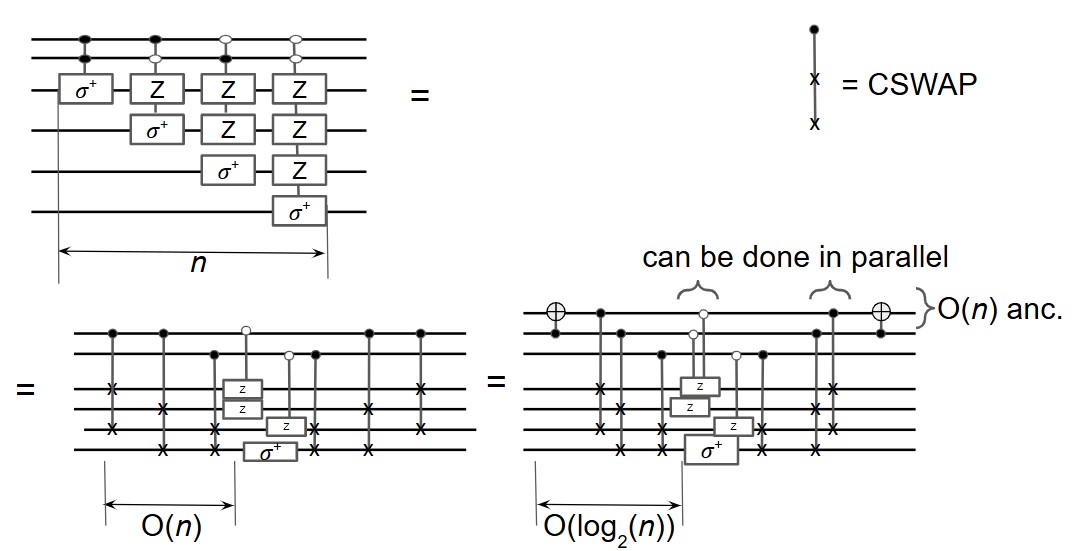}
    \caption{Three equivalent ways of preparing $\sum_{\mu=1}^M \ket{\mu}\otimes c_\mu^\dag \ket{\psi}$, where $c_\mu^\dag$ are creation operators on site $\mu$. In this example $M=4$. The left/middle figures are implemented with depth $O(M)$ and ancilla cost $O(\log M)$, whereas the rightmost has an ancilla cost $O(M)$ and a depth $O(\log M)$. The curly braces in this figure indicate the CSWAPs and CZs can be implemented in parallel. The CNOT comes from the GHZ preparation circuit (see main text).}
    \label{fig:ferm}
\end{figure}

We will utilize the same structure for the translation trick: controlled on a single bit of $\mu$, there are two sides with a single application of $\sigma^\pm$ respectively and $Z$'s that have to be conditionally applied on up to 19 qubits. The interplay of $Z$ and SWAPs is nontrivial and is illustrated in Fig. \ref{fig:ferm}. A single such circuit on 20 qubits will have $\sim$40 CSWAPs,  19 CZ, and a SWAP for $\sigma^\pm$. We need to control it on the additional two qubits: one for the left-right side and another for the control of CU. There are two of those circuits, and 180 C$^2$Z. The counts for CU are:
\begin{equation}
    \text{T-count: } 1200, \text{ depth: }32
\end{equation}
The leading contribution to the number of qubits is $3\cdot 200$ for the system and the ancillae controlling SWAPs and $Z$.

\subsection{Cost of time evolution}

The T-counts, T-depth, and depth of $CU$ for the two applications can be used to compute parallelism factor $k = T$-count$/5\cdot$depth (the factor 5 is due to the depth of teleportation needed for implementing possibly spatially non-local CNOTs in a local architecture), as summarized in a table:

\begin{center}
\begin{tabular}{ |c|c|c| } 
 \hline
$CU$ & Hubbard & Ca$_3$Co$_2$O$_6$ \\ 
\hline
 T-count & $1,200$ & $185,000$ \\ 
 qubits & $600$ & $4000$ \\ 
 depth & $32$ & $23,000$ \\
 $k$ & 7.5& 1.6 \\
 \hline
\end{tabular}
\end{center}

While for Ca$_3$Co$_2$O$_6$ the Hamiltonian is much simpler than the open system generator, and the cost of its block encoding $V$ can be neglected, the opposite is the case for the Hubbard model: while the open system operators are concentrated at the boundaries, the Hamiltonian terms cover the entire lattice. 
To obtain the counts of the Hubbard Hamiltonian block encoding $V$, we utilize the recently created PyLIQTR library.  A brief summary of our numerical experiments is in App. \ref{app:plqtr}. The resulting counts are:
\begin{equation}
    \text{T-count: } 14840;\quad  k \sim 1
\end{equation}
This removes any possible gains of parallelism. It may be possible to recompile the Hamiltonian block encoding using the translation trick and recover the parallelism, which we leave for future work. As there are only 200 qubits, $CV$ can be implemented via CSWAPs of all of them at the cost of $1600$ T gates. The total count for $CV$ and $CU$ is: 
\begin{equation}
    \text{T-count: } 18000,\quad  k \sim 1
\end{equation}

Besides $U$ and $V$, the algorithm in Theorem III.2 of \cite{chen_quantum_2023} also uses other gates. We will assume that the additional 2-qubit gate count $\sim t \text{log}M\text{log} t/\epsilon$ stated in \cite{chen_quantum_2023} can be roughly interpreted as the T-gate count, as it is subleading anyways.

Recall that the algorithm in Theorem III.2 of \cite{chen_quantum_2023} uses the following number of consecutive calls to $CU$ and $CV$:
\begin{equation}
    O\left((\alpha t + 1) \frac{\text{log}((\alpha t + 1)/\epsilon) }{\text{loglog}((\alpha t + 1)/\epsilon)}\right)
\end{equation}
where $\alpha$ is the time upscaling factor discussed above ($\alpha = 1100$ and $6500$ for the two cases we consider), and $t$ the desired simulation time. To estimate the various counts of the whole algorithm, we multiply it by the corresponding count of the block encoding $CU$ and drop the constant factor hidden in the big-O notation:
\begin{equation}
    \text{alg. count} =\alpha t \frac{\text{log}(\alpha t /\epsilon) }{\text{loglog}(\alpha t /\epsilon)} \times \text{ count for }U\ ,
\end{equation}
For Hubbard, $t=36$ and $\epsilon=0.1$, while for Ca$_3$Co$_2$O$_6$, we need to makes $n_{s} =10$ steps of $t=10^{11}$, each with precision $\epsilon=0.01$. Plugging in these numbers, we get: 
\begin{center}
\begin{tabular}{ |c|c|c| } 
 \hline
~ & Hubbard & Ca$_3$Co$_2$O$_6$ \\ 
\hline
$n_s\alpha t \frac{\text{log}(\alpha t /\epsilon) }{\text{loglog}(\alpha t /\epsilon)}$ & $1.7 \cdot 10^5$ & $6.3 \cdot10^{16}$\\
 T-count & $3.3\cdot 10^{9}$ & $1.4\cdot 10^{22}$ 
 \\ k & 1 & 1.6 \\
 qubits & $600$ &$4000$\\
 \hline
\end{tabular}
\end{center}
Note that the Trotter open system time evolution algorithm analyzed exactly in App \ref{app:ex} yields $1.27\cdot 10^{42}$ T-gates for Ca$_3$Co$_2$O$_6$.

\subsection{Physical}
To estimate the physical requirements, we utilize the Microsoft Azure Resource Estimator introduced in \cite{beverland2022assessing}. We note that recent papers such as \cite{gidney2024magic,hirano2024leveraging} may significantly improve those estimates, especially under the assumption of $10^{-4}$ noise. For our estimates, we use the default $10^{-3}$ noise on superconducting qubits and the {\texttt{AccountForEstimates}} function with the above numbers. For Ca$_3$Co$_2$O$_6$, $2.1\cdot 10^{22}$ T-gates is too large, and the algorithm fails, which is why we extrapolate for 4000 qubits from the functional dependence in the range of $10^3 .. 10^{18}$ T-gates. We then divide by $k=1.6$ to estimate the runtime of the complication that does not get rid of Cliffords but takes advantage of the parallelism. The results are:

 \begin{center}
\begin{tabular}{ |c|c|c| } 
 \hline
 ~ & Hubbard & Ca$_3$Co$_2$O$_6$ \\ 
\hline
  runtime (s)  & 38,000 &  $2.2 \cdot 10^{17}$  \\ 
  ~ & ($11$ hours) & (7b years)  \\
  physical qubits & 2.9M& 46M\\
  factory qubits & 0.8M &4M\\
 \hline
\end{tabular}
\end{center}

In the MIT material search, we need to execute $10^3$ runs for each model of the $10^3$ models. Even if we reduce their complexity to the same value as the Hubbard model above, one may still need to execute them in parallel on multiple quantum computers or take years as experts also need time to come up with each next model. We expect only the hardest models, say $1\%$ of 1000 suggested ones, to be sent to the quantum computer, while others would be investigated by classical methods or experimentally.

\begin{figure}[t]
\centering
\includegraphics[width=\columnwidth]{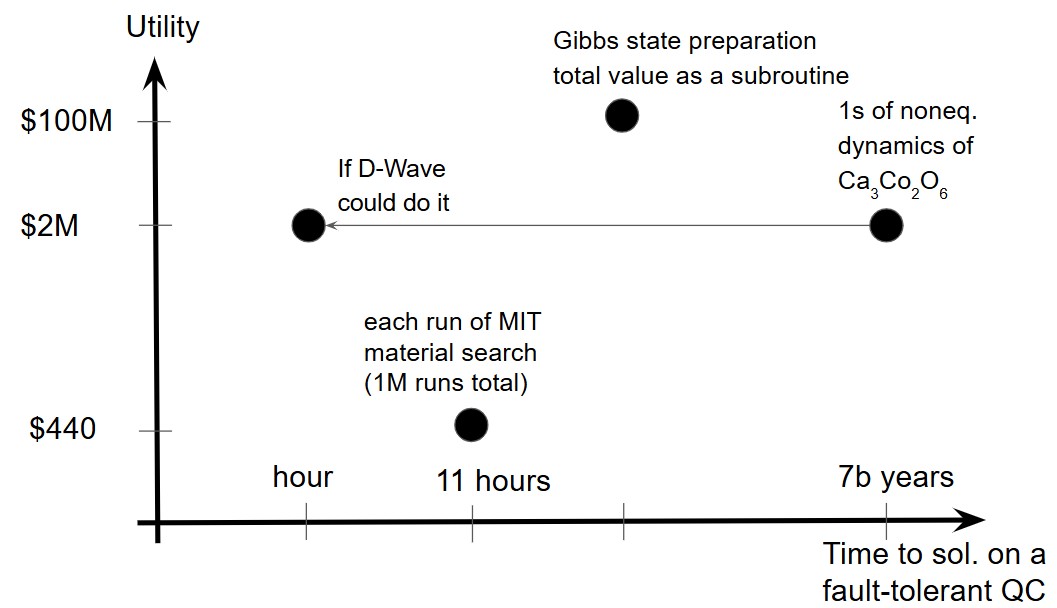}
\caption{We plot the three discussed applications of open system quantum simulation. It should be assumed that the points have 1-2 orders of magnitude error bars}
\label{fig:jp}
\end{figure}

We collect the results of our estimates on the resource estimate vs. utility plot in Fig. \ref{fig:jp}. For completeness, we also indicate where analog quantum simulators \cite{king2021quantum} may lie. The correct non-equilibrium behavior was observed on D-Wave, yet there are many hardware constraints preventing a true {\emph{ab initio}} simulation. While the runtime of the universal quantum computer appears discouraging, we note that our estimate is neither overly optimistic nor overly pessimistic. The numbers may improve with the invention of a qubit hardware that is faster or more noise-resilient than the current superconducting qubit technology with $10^{-3}$ error rate and $300$ns code cycle that we used. We also expect algorithmic improvements as the open system simulation algorithms are in the early stages of development. The fault-tolerant architecture itself may change as the first experimental tests of it enable its optimization beyond the conventional T-state distillation approach. On the other hand, the factors in big-O notation and the subleading terms, as well as the contribution of the Hamiltonian dynamics we dropped in this estimate would likely increase our numbers by several orders of magnitude. The hardware may also experience new limitations as it is scaled up to tens of millions of qubits, which are unseen in the current generation of devices. We hope our investigation pointed out the urgent open questions in the field and provided a more tempered expectation for a broader community. 

\section{Acknowledgments}
We thank Tameem Albash and Joshua Job for contributing to an early version of the draft. 
We thank Kaitlyn Morrell, Kevin Obenland, and the rest of the team working on pyLIQTR for implementing new Hamiltonian models for us \cite{pyLIQTR}.

This material is based upon work supported by the Defense Advanced Research Projects Agency (DARPA) under Contract No. HR001122C0063. J.M. is thankful for support from NASA Academic Mission Services, Contract No. NNA16BD14C. N.A. is a KBR employee working under Prime Contract No. 80ARC020D0010 with the NASA Ames Research Center. The United States Government retains, and by accepting the article for publication, the publisher acknowledges that the United States Government retains, a nonexclusive, paid-up, irrevocable, worldwide license to publish or reproduce the published form of this work or allow others to do so, for United States Government purposes. All material, except scientific articles or papers published in scientific journals, must, in addition to any notices or disclaimers by the Contractor, also contain the following disclaimer: Any opinions, findings and conclusions or recommendations expressed in this material are those of the author(s) and do not necessarily reflect the views of the Defense Advanced Research Projects Agency (DARPA).

\bibliographystyle{apsrev4-2}
\bibliography{refs}

\appendix

\section{A sketch of an economic analysis of utility}
\label{app:eco}

We note that technology is definitely not enough to create a product; one also needs people and assets, as well as various other costs. While a company cannot fill the market in question without the corresponding technology, it would not be accurate to count the entire revenue as the utility of the technology. The founder who discovers the technology either invests his own money or seeks venture funding by selling shares of the company when the company is created. Only with that money can the company fill the market. Since there are various monetary figures at various stages of this process, and the revenue itself grows with time, we need a standardized method to reduce it all to a single number - utility. Note that the first stage of venture funding usually values the company relatively cheap, no matter how good the underlying technology is. We see that as the tradition of venture capital and not as the true value of the corresponding technology. We believe that the true value of technologies shows itself in acquisition events, where a big company acquires promising ideas before they become competitors. We also note that for utility estimates, the total revenue is less important than the relative increase with or without the technology. Ideally, that can be computed from the business model and an estimate of how much the technology improves the product. That being said, none of those numbers can easily be used in a unified framework, as they are rarely available. We condense all of these considerations into a factor between 0.1 and 10 that is used in our methodology.
The factors given as examples in the main text are $1/2$ for nitrogen fixation alternatives and $10$ for thermoelectric devices.

A more detailed estimate of the factor $1/2$ would be as follows: first, we assume a simple model of seller optimizing profit, and the number of sales n depending linearly on price as $n = n_0 -ep$, where e is the elasticity of demand. The result of profit optimization for production costs $c1$ and $c2$ is that the difference in revenue is $r_2-r_1= \frac12 e(c_1 - c_2)$. As the fertilizer market is tied with the population and the arable land, we can estimate elasticity as the remaining need for fertilizer divided by the current price since, at zero price, only the amount needed for current land and population would be purchased.  The amount of fertilizer needed would remain approximately the same over a 10-year span. Finally, we estimate lower production costs $c_2$ from the reduction in energy consumption to fix nitrogen. Various methods have $O(1)$ difference in energy cost per unit of mass of nitrogen fixed, so we can expect a novel method to have of order 50\% improvement. If it is only a 1\% improvement, the estimate can be updated by directly multiplying the factors.  Assuming the 50\% energy savings, prices close to the production cost, and elasticity $\sim n_0/$current price, this estimate results in a factor $1/2$ improvement in revenue for the new nitrogen fixation technology.

In contrast, a slightly better product requiring a lot of investment into switching the manufacturing process completely would have two competing effects: rising production costs would lower the revenue, yet the improvement in product quality would increase the revenue, though the effect may be quite complex to formulate. For example, consider thermoelectric devices. While they are sold commercially today, the market is small, mostly consisting of niche cooling applications. The main application of recycling heat is mostly unrealized. It is believed that improvements in efficiency (essentially, the percentage of heat recycled) would make it viable and result in mass adoption. More specifically, the efficiency, together with other figures of merit involving lifespan, production, and maintenance cost of the thermoelectric device, determines the viability for various use cases. The demand in each use case would discontinuously jump from zero to mass adoption, but if one adds these jumps for all use cases, one could approximate the number of sales as linear in the figures of merit. Using the formalism above, that would be $n_0 =a  ZT$. The change in revenue would then be $en_0a(ZT_2-ZT_1)$. If one expects a tenfold increase in the number of sales, that would roughly translate to a tenfold increase in revenue, assuming other coefficients are $O(1)$. That illustrates that technologies with a small existing market can have much higher utility than the current or projected market size suggests. This applies only to conventionally impossible tasks, e.g., profitable heat recycling is essentially impossible with the current technology; therefore, the existing market is not a good estimate for the potential market. For most technologies, however, the task is already accomplished by the older generation of technology, and the shift in demand is not as dramatic, no matter how good the new generation of technology is.

\section{Planted solution problems}
\label{sec:app}
To accompany our application instances where the answers are unknown, here we suggest some planted solution instances with known answers but no direct application relevance, including the open system and the closed system versions. In the latter, we note the possibility of obfuscation that would prevent discovering the underlying planted solution.

\subsection{Closed system: Integrable models and T-gates}

Some prior work on breaking the type of obfuscation considered below can be found in \cite{pozsgay2024quantum} and references therein. The theorems those references prove concerned themselves with the situation when the solution is known via some free fermion ansatz for any value of coefficients of the Hamiltonian terms. Then, given just the terms, one can always recover the ansatz. In the construction below, the model is solvable only for fine-tuned values of coefficients, and these theorems are not applicable.

The problem is formulated as follows.

{\noindent Input:} 
\begin{itemize}
    \item A superposition of stabilizer states defined on $n$ qubits.  Instead of specifying the state, the operators (in our case, Pauli operators) that stabilize the state are provided.
    \item A Hamiltonian on n qubits expressed as a linear combination of Pauli operators.
    \item A total simulation time $T$.
    \item List of observable(s) $A$.
    \item Desired relative error $\epsilon$.
\end{itemize}
Output:
\begin{itemize}
    \item Expectation value of observable(s) $A$ at time $T$ to a relative error $\epsilon$.
\end{itemize}
In order to ensure that we have a known solution, we use an integrable model as our starting Hamiltonian. For example, we can consider the 1-dimensional transverse field Ising model (open boundary conditions):
\begin{equation}
   H = \sum_{i=1}^{n-1}J_i Z_i Z_{i+1} + \sum_{i=1} h_i X_i  
\end{equation}

For concreteness we can take $J_i =-1$ and $h_i =-1$.  For our initial state we can take the state $|++\dots +\rangle$ where $ X |+\rangle=|+\rangle$ and $X |-\rangle=-|-\rangle$.  This state is stabilized by the set ${X_1,X_2,...,X_n}$. We can define a set of observables $A_{ij}=Z_iZ_j$ such that we can calculate the expectation value of these operators at time T efficiently by mapping the problem to a free fermion problem.

Our aim now is to obfuscate the initial state, the Hamiltonian, and the observables. Applying a random Clifford circuit to all three is not enough because it preserves the dimension of the dynamical Lie Algebra, which only scales polynomially with system size for this problem. We therefore propose to first apply a small number of T-gates followed by a random Clifford circuit.  Our initial analysis indicates that even applying a single T-gate is enough to ensure that the dynamical Lie Algebra scales prohibitively. Some issues we are aware of:
\begin{itemize}
    \item The weight of the terms in the obfuscated problem is larger than the weight of the operators in our application problem classes, which naively suggests it is a harder dynamics problem. Generating Clifford circuits that maintain a weight of 2 and obfuscate the problem, if possible, will require a more sophisticated strategy.

    \item We are not currently aware of an algorithm that can efficiently identify the integrable model that we have obfuscated.  The algorithm in \cite{gunderman2023occam} is only able to identify the $Z_2$ symmetry of the model but not identify the original Hamiltonian. It is, however, possible that the algorithm that breaks our obfuscation would be discovered.
\end{itemize}
 
\subsection{Open system}

In this section, we outline a toy problem with a known solution that can be used to test an open system simulation of any size. We suggest reproducing the results of \cite{prosen2008third}. The specific details are below:

\begin{equation}
   H = \sum_{i=1}^{n-1}J_i \sigma^z_i \sigma^z_{i+1} + \sum_{i=1} h_i \sigma^x_i  
\end{equation}
The values $J_i = J =1.5$ and $h_i = h =1$ are chosen.
The system is also attached to two baths of different characteristics, given by the following Lindblad generators:

\begin{align}
    L_1 = \frac12 \sqrt{\Gamma_L^-} \sigma_1^-, \quad L_2 = \frac12 \sqrt{\Gamma_L^+} \sigma_1^+,\\ L_3= \frac12 \sqrt{\Gamma_R^-} \sigma_n^-, \quad L_4 = \frac12 \sqrt{\Gamma_R^+} \sigma_n^+\ .
\end{align}
The rates are $\Gamma_L^- =1, ~\Gamma_L^+=0.6, ~\Gamma_R^-=1, ~\Gamma_R^+=0.3$. The task is to simulate the open system evolution of this system.

The initial state is all $|0\rangle$ state, and the evolution time needs to be larger than the inverse of the gap of the Liovillian above, which is known to be $\approx 10.9n^{-3}$.

The observables of interest are energy current and spin current. If the chain Hamiltonian is presented as a sum of local terms $H=\sum_m H_m$, the energy current is $Q_m = i [H_m, H_{m+1}]$, and the spin current $S_m = i[\sigma^z_m,H_m+H_{m-1}]$. The average energy and spin current are given as follows 
\begin{equation}
    S= \frac{1}{n-1}\sum_m S_m
\end{equation}
Their target values for different system sizes and chain locations are given in \cite{prosen2008third}. The energy current saturates at $0.35$, while the spin current is at $0.12$, with a uniform profile in the bulk of the system, corresponding to the ballistic transport.

While these models, as well as a large collection of other models that can be mapped to free fermions, have an efficient classical solution, it is a fair estimate of the resources required to solve actual application problems. The main differences in applications are higher dimensionality, a larger number of fermion bands, and the presence of interaction terms quadratic in fermionic operators. Our application instances range between 2000 and 15000 problem qubits, with 2000 to 3000 Lindblad terms. A successful quantum simulation of the $n=15000$ qubit benchmark would indicate that the benchmarked device has sufficient capabilities to contribute to application scenarios of MagLab and MIT search as well.

\section{Known gate compilations}
\label{app:howtocount}
\begin{figure}[t]
\centering
\includegraphics[width=\columnwidth]{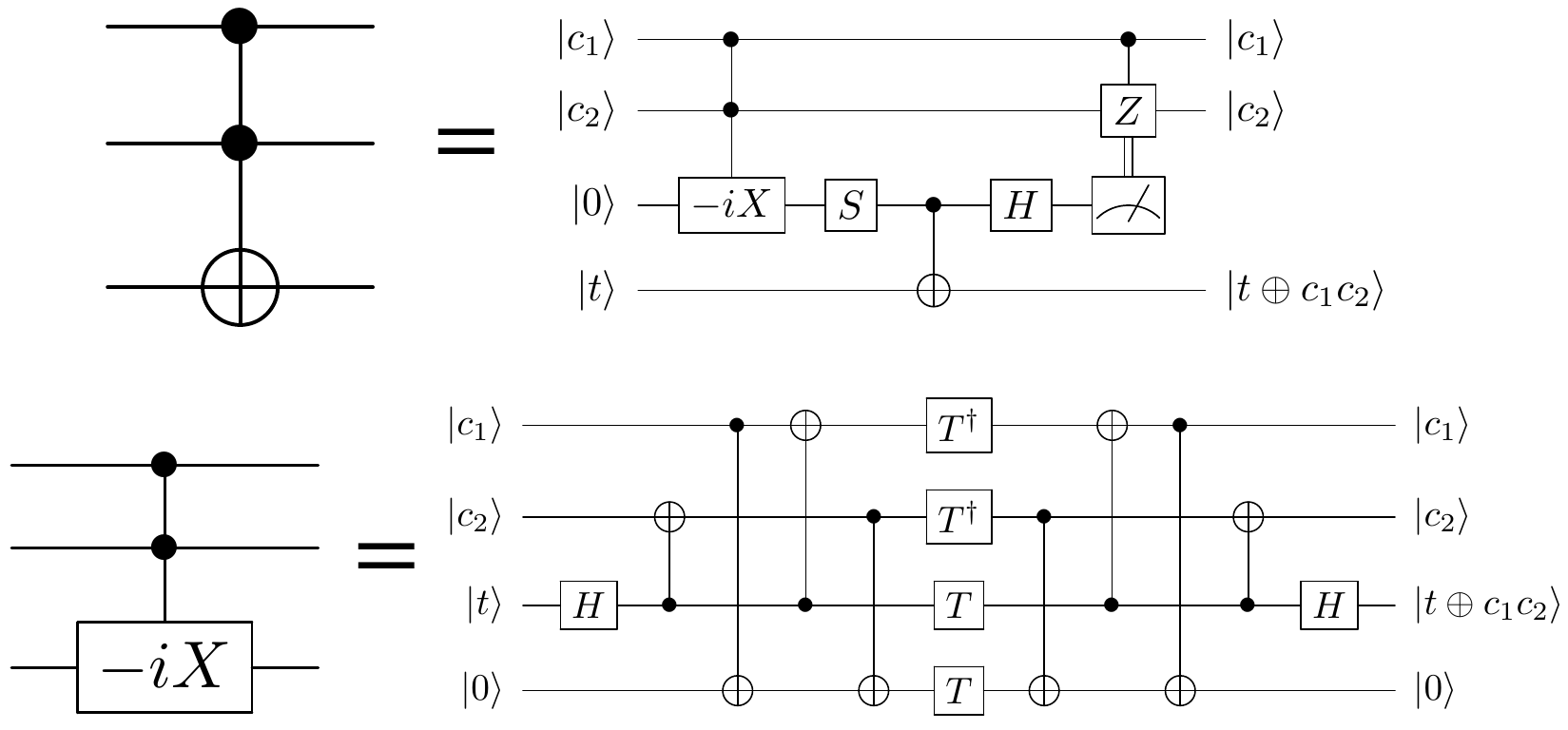}
\caption{An implementation of a Toffoli with 4 T-gates, from Ref.~\cite{Allen_toffoli}, using results from \cite{Jones-overhead} and \cite{Selinger-depth-one}.} 
\label{fig:4T}
\end{figure}

In this section we collect the gate compilations we chose to use in the main text, along with references and discussion. Gate compilation is a problem of compiling a wide range of gates appearing in quantum algorithms into a specific gateset. Our algorithms are written for future fault-tolerant quantum computers. The resource estimates for those depend heavily on the T-state footprint. For example, the tool \cite{beverland2022assessing} allows for an estimation of physical resources provided the following circuit parameters. First, the circuit is arranged into Clifford (and measurements) and non-Clifford diagonal layers (CCZ, T, and arbitrary rotation $R$ gates). The estimator then asks for the counts of T-gates, $CCZ$-gates, and arbitrary single qubit rotations $R$, as well as rotation depth (number of layers containing $R$) and measurement count. The circuit is then compiled in such a way that the cost of Cliffords is completely removed, while each non-Clifford gate is now performed via operations on up to all qubits, so any parallelism is lost, and non-Cliffords must be applied sequentially. A traditional compilation, in contrast, still performs both Cliffords and T-gates, so it can take advantage of the parallelism, but the total depth is can be much higher due to the Cliffords. Alternative compilation tools such as Bench-Q \cite{benchq} take some advantage of the parallelism but do not go to the extreme space-time tradeoff of consuming T-state every logical cycle. For our estimates, we always try to optimize depth and T-count, possibly at the cost of introducing extra ancillae and operations conditioned on the measurement of those ancillae. With these capabilities, a $CCZ$-gate can be performed using 4 T-gates and T-depth 1, total depth 11 (Fig. \ref{fig:4T}). 4 T-gates is the conversion used by the tool \cite{beverland2022assessing} as well.

We now discuss controlled gates $CU$. Let unitary $U$ be given to us by its decomposition into a product of gates $U =\prod_i U_i$. In case these gates are just Pauli's and CNOTs, a naive approach $CU =\prod_i CU_i$ may be the best option (optimal T-count and T-depth). But even for Pauli's and CNOTs, if the number of CNOTs is larger than twice the support $n$ of $U$, it is more optimal to add $n$ ancillae, do CSWAPs between the original $n$ qubits and the ancillae, and then perform $U$ on the ancillae and CSWAP back. That is because CSWAP is a Toffoli and two CNOTs, so we get a better Toffoli count. Another trick is that if $U$ contains a circuit that undoes itself, we do not need to add controls on it, only to what's in between the two sides of it:
\begin{equation}
    U = U_1 U_2 U_1^\dag, \quad CU = U_1 CU_2 U_1^\dag \label{propCU}
\end{equation}

\begin{figure}[t]
\centering
\includegraphics[width=0.5\columnwidth]{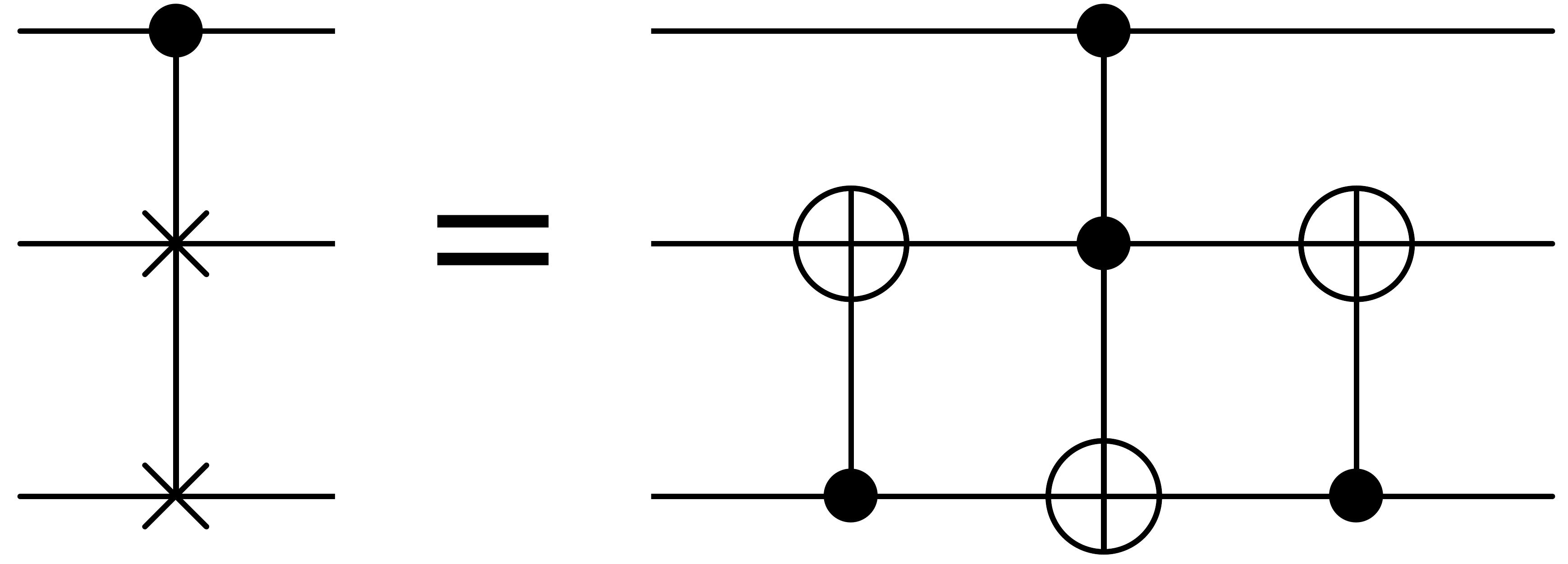}
\caption{A an implementation of a controlled swap using a CCX gate \cite{Crooks-gates-notes}.} 
\label{fig:tof}
\end{figure}

The counts for CSWAP itself are 4 T-gates and total gate depth $13$, applying the Toffoli decomposition to (\ref{fig:tof}).

We note that though our operation $U$ is unitary, the way it is implemented may contain an introduction and measurements of an ancilla and classical operations conditioned on the measurement outcome. In that case $CU$ may be given by a circuit where all the unitary operations are conditioned on the new qubit while measurements still happen. The CSWAP-based implementation remains unchanged.

\begin{figure}[t]
\centering
\includegraphics[width=\columnwidth]{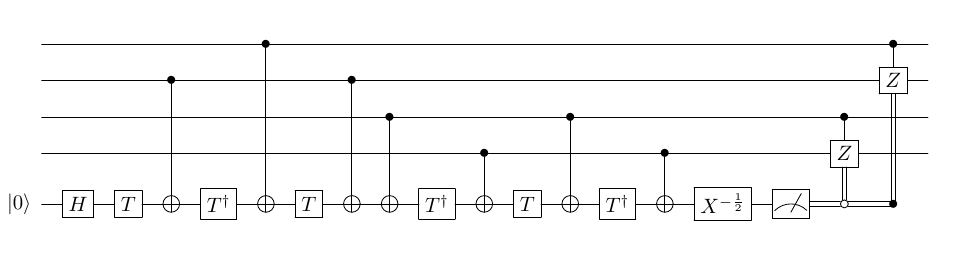}
\caption{An implementation of a $C^3X$ with 6 T-gates, from Ref.~\cite{gidney_cccz_2021}.} 
\label{fig:C3X}
\end{figure}

Multiple-controlled gates $C^nU$ almost always benefit from introducing one extra ancilla, performing $C^nX$ on it, and then $CU$ and undoing $C^nX$. Only if $U$ is a single gate or something that itself may be compiled, might there be benefits in compiling the controls differently. The gate $C^nX$ can be compiled in various ways. A nearly optimal T-count is $4n-4$, which takes $16$log$_2n +12$ depth \cite{he2017decompositions}. There is a slight improvement (Fig. \ref{fig:C3X}) for $n=3$ (controlled Toffoli) with 6 T-gates (instead of 8), and depth 16 (instead of 37).

We do not intend to keep track of all the details for the depth count in the leading order estimates in the main text. For example, $C^nX$ and $C^nZ$ will be counted to have the same depth, even though they may differ by two layers of Hadamards. 

We now sum up the tricks involved in adding some number of controls $C^mU$ to a gate $U$ on $n_q$ qubits. We want the individual gates $U_j$ in $U = \prod_j U_j$ to be split in two groups: to be controlled group $j\in\mathcal{C}$ and pairs of gates undoing each other $i,i' \in \mathcal{U}$. For the gates considered for the leading order estimates of the algorithms in this work, the controlled group will only contain $J$ gates of the form $C^{n_j}X$ for $n_j> 0$, and $P$ Paulis, that for the discussion below may be thought of as $C^0X$ up to undone pairs of Hadamards. There are three ways to implement $C^mU$:
\begin{enumerate}
    \item Every $C^nX$ in $\mathcal{C}$ is replaced by $C^{n+m}X$
    \item An ancilla is prepared and undone by two applications of $C^mX$, and then every $C^nX$ in $\mathcal{C}$ is replaced by $C^{n+1}X$
    \item An ancilla is prepared and undone by two applications of $C^mX$, duplicated $n_q$ times, and $n_q$ CSWAPS are performed, then $U$ is applied normally.
\end{enumerate}
The expressions (in the limit of large $m, n_j$) for the corresponding additional costs of $C^mU$ compared to $U$ are:
 \begin{center}
\begin{tabular}{ |c|c|c| } 
 \hline
method &+$n_q$ & +T count \\ 
\hline
1.&$n_r $ & $4 (J m +P(m-1)) $\\ 
2.&$n_r$& $8(m-1) + 4J$  \\
3.&  $\sim 2n_q$ & $8(m-1) + 8n_q $\\
 \hline
\end{tabular}
\end{center}

 \begin{center}
\begin{tabular}{ |c|c| } 
 \hline
method & +depth \\ 
\hline
1.& $ 16\sum_j' \text{log}_2\frac{n_j+m}{n_j} + 4P(4 \text{log}_2 m -3)  +r$\\ 2. & $16\sum_j' \text{log}_2\frac{n_j+1}{n_j} +8(4\text{log}_2 m -3)+r$ \\3.&  $8(4\text{log}_2 m -3) +2 \text{log}_2n_q +26$\\
 \hline
\end{tabular}
\end{center}
Here $\sum_j'$ sums over $j$ with $n_j\ne 0$. Each of these methods will be optimal for some values of $n_q, J,P, m$, and $r,n_r$ stand for the cost incurred by breaking the parallelism in the control group (for the strategy of repeating the control while preserving the parallelism, $r \leq 2 \text{log}_2n_q, n_r \leq n_q$, $\times m$ for the first method). For example, for $C^2U_\pm$ from Sec. (\ref{sec:1l}) $n_q\sim P \sim 100, J=1, m=2$, the best T-counts are obtained by the second method. Repeating the control hurts depth way less than applying CZ in sequence, so we chose the former. $r$ is either $0$ or $r \sim \text{log}_2 180 \sim 8$ for the appropriate side. Moreover, the creation and undoing of the repeated register can be applied in parallel with the depth 14 CSWAP, which makes the depth of the corresponding sides $14$ and $16$. This results in $C^2U_\pm$ counts used in the main text:
\begin{equation}
  \text{Max. qubits: }401,  \text{  T-count: }12, \text{ Av. depth: }37
\end{equation}
For the neighborhood dependent rates $C^kCU_\mu$ from Sec. (\ref{sec:1l}), only three gates are not undone in $U_\mu$: $C^6X, CX, X$, the latter two applied on the same qubits in sequence. The values are $J=2, P=1, n_q = 17, m=k+1 = 1 +$log$_239M$. The second method is again optimal with the counts:
\begin{equation}
  \text{+qubits: }1,  \text{  +T-count: }8(k+1), \text{  +depth: }4(8\text{log}_2(k+1) -5)
\end{equation}
We also note that the application of CSWAPs or $C^nX$ that is later undone may allow to neglect certain phases or even use non-Clifford gates only on one side of the compute-uncompute arrangement, as done for $C^nX$ in Fig. 5 of \cite{gidney2018halving}. According to them, the T cost of two $C^2X$ in the circuit $C^2X \cdot U  \cdot C^2X$, which should usually be 8 T-gates, can be halved if the circuit possesses the following property:
\begin{equation}
    C^2X \cdot U  \cdot C^2X = \langle 0 |_a C^2XX_a \cdot (U \otimes {\mathds{1}})  \cdot C^2XX_a| 0\rangle_a
\end{equation}

With that in mind, the first gate we will consider is a CCSWAP. As CSWAP is a Toffoli between 2 CNOTs, those will cancel without it. By Eq. (\ref{propCU}), we can just add another control obtaining $C^3X$. The resulting count is 6 T-gates, depth 18, and one ancilla (using Fig. \ref{fig:C3X}). We do not apply this trick to the cost estimates of the third method, leaving it to future work.

\begin{figure}[t]
\centering
\includegraphics[width=\columnwidth]{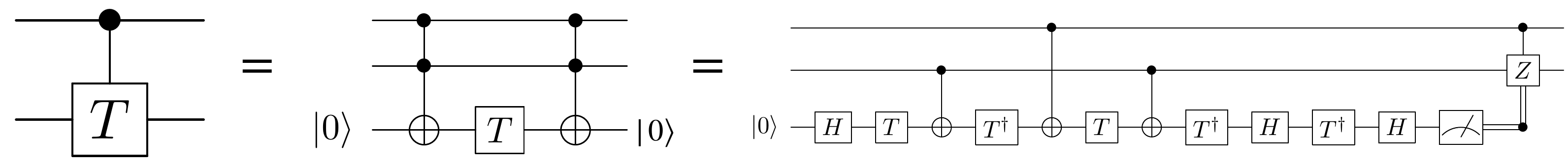}
\caption{A an implementation of a controlled T-gate, from \cite{gidney_CT}.} 
\label{fig:CT}
\end{figure}

One more gate we will need is $CT$. Its T count turns out to be $5$ and depth 13 (see Fig. \ref{fig:CT}). 

\section{Average MIT material search block encoding}
\label{app:av1019}
Here, we do a preliminary resource estimate for the thermal generators $L_\mu^{\text{therm}}$ proposed in \cite{chen_quantum_2023}. 
One starts with an environment interaction operator $ c_0$ or $c_0^\dag$ that describes some orbital in the chosen description of FeMoCo. The operators $L_\mu^{\text{therm}}$ are constructed from time integrals of 
 the Heisenberg operators $c_0(t), c^{\dag}_0(t)$. We note that in our non-interacting lattice, $c_0(t), c^{\dag}_0(t)$ are evolved with the local FeMoCo Hamiltonian, which is meant to represent a more sophisticated heuristic taking advantage of the approximate locality of $L_\mu^{\text{therm}}$. that allows one to truncate the time integral to $\int_0^{2^nt_0}$ and discretize it in steps of $t_0$, where the parameters $n$ and $t_0$ are provided in Corollaries C.1, C.2 of \cite{chen_quantum_2023}. We can loosely interpret the bounds there for $\beta, \tau_B, \tau_{SB} \sim 1$ as:
\begin{equation}
    n \sim \text{log}_2 (\|H\| |A|/\epsilon_{L}), \quad t_0 \sim 1/\sqrt{2^n}
\end{equation}
Here $|A|=20$ is the total number of environment interaction operators, $\epsilon_L$ is such that for our experiment time $T$ the value $\epsilon_L T \sim \epsilon$, the error tolerance.

\begin{figure}[t]
\centering
\includegraphics[width=\columnwidth]{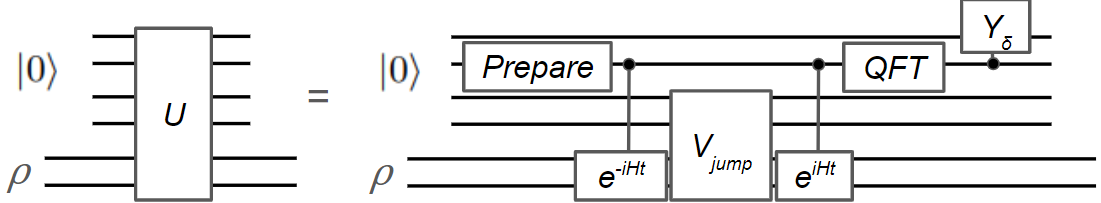}
\caption{A circuit for the block-encoding of a thermal Lindbladian $U$ used in \cite{chen_quantum_2023}. The gate $Y_\delta= e^{-iY\text{arcsin}\sqrt{\delta}}$ and here $\delta=1-\gamma(\omega)$ is taken conditionally on the frequency of each contribution to the Lindblad operator}
\label{fig:CGcir}
\end{figure}

The block encoding $U$ used for efficiently implementing $L_\mu^{\text{therm}}$ used for Gibbs state preparation is developed in \cite{chen_quantum_2023}. Those generators rely on the ability to implement the controlled closed system evolution, as well as the custom gates {\textit{Prepare}}, $V_{jump}$ that, together with quantum Fourier transform (QFT) allow to integrate that closed system evolution. We do not go into the detail of that construction, providing a high-level overview of the circuit in Fig. \ref{fig:CGcir}.

The total cost is given in Lemma I.1 of \cite{chen_quantum_2023}: the maximum closed system controlled simulation time is $T_{CG}$, the internal parameter of $L_\mu^{\text{therm}}$, and besides $n$ system qubits extra ancillas logarithmic in the number of jump operators $A$ and in the Fourier discretization grid size $N$ are used. The internal simulation time $T_{CG}$ and grid size $N$ are specified in \cite{chen_quantum_2023}. We note that for preparing the equilibrium state (in contrast to the open system evolution), this is not the optimal algorithm, and the authors of \cite{chen_quantum_2023} present further refinements. 

For the average MIT material search problem, the cost of Lindblad generators in Fig. \ref{fig:CGcir} will be dominated by the controlled $\sum_{t}|t\rangle \langle t| \otimes e^{iHt}$ evolution with the single-site Hamiltonian. The time is discretized in $2^n$ steps of $t_0$ in the notation of \cite{chen_quantum_2023}. The existing QSP approaches rely on interleaving the Hamiltonian block encoding $V, ~\langle 0|^{\otimes b} V |0\rangle^{\otimes b} $ and the controlled rotations $\Pi e^{i\phi} +(1-\Pi), ~\Pi = (|0\rangle\langle 0|)^{\otimes b}$ to create polynomials approximating functions such as sgn$H$, $\cos 
Ht, \sin Ht$ from which $e^{iHt}$ can be constructed (for technical reasons, the coherent construction \cite{martyn2023efficient} requires those intermediate steps). The number of steps and the phases $\phi$ are determined by the target precision and the desired evolution time $t$ in a classical efficient procedure \cite{low2016methodology}. The complexity and compilation of that procedure to a quantum computer is only known to be polynomial. Alternatively, one can use linear in size $n$ of the $|t\rangle$ register number of usual time-evolutions for times $2^{n-1}t_0, \dots t_0$, each controlled on the corresponding bit of the time register. We will need $n$ single-bit-controlled time evolutions. Either way, the leading contribution is likely to come from the Hamiltonian block encoding used a logarithmic number of times.  A total time is of order $2^n t_0 \sim \sqrt{2^n} \sim (\|H\| |A|/\epsilon_{L})^p$, where the power $p$ can be made a small constant for the right construction of the bath parameters. We will set $p=1$ in our estimate, resulting in a controlled-time Heisenberg operator, and thus the block-encoding cost of a single Lindblad generator, to be estimated as:
\begin{equation}
    \lambda^2\sqrt{2^n}\times (T\text{ cost of }V) \sim \frac{\lambda^3\cdot 40\cdot T_{\text{Av.}} \cdot 17000}{\epsilon}\sim10^{19}
\end{equation}
This large number indicates that using the Landauer approach is unrealistic for practical ab initio simulations. Note that the chosen patch size $10\times10$ had negligible effect, and most of the cost comes from the complicated electronic structure Hamiltonian of a single lattice site and from the chosen method to create sources and sinks that are guaranteed to have a certain temperature and chemical potential. It would be much more efficient to approach this in a heuristic manner by trying various Lindblad generators and measuring the steady state temperatures and chemical potentials they induce. Resource estimation for such heuristics is much harder than for guaranteed approaches, so we limit it to our Hubbard model example.

\section{Numerical analysis of the closed system Hubbard model embedding}
\label{app:plqtr}
We consider the MIT Hamiltonian Eq.~\eqref{eq:HM}, which is a next-nearest-neighbour Fermi-Hubbard model. It is defined on a square lattice with hopping terms $t$, with additional couplings in the diagonal directions of strength $t'=0.15t$. The on-site density term has strength $U=20t/3$.

We use \textsc{pyLIQTR} \cite{pyLIQTR} to generate resource estimates for the block encoding for a $10\times 10$ lattice, with open boundary conditions, for $t=1$. The model is encoded to qubits via the Jordan-Wigner transformation. We consider the standard Pauli linear combination of unitaries (LCU) encoding. Assuming a target precision of $10^{-11}$ for rotation gates (our resource estimate will return $3.3\cdot 10^9$ T-gates at precision $0.1$, so this is self-consistent), the logical complexity for the block encoding circuit is: 211 qubits and 471032 T-gates. 
Using the ``linear-T" (LT) encoding of \cite{babbush2018encoding}, the cost can be reduced to just 14840 T gates (T-depth 5997), with a similar number of qubits.

Additionally, with \textsc{Qualtran} \cite{harrigan2024expressing}, we can compile this to obtain physical resource estimates. With a distance 21 surface code, it is possible to achieve logical error $\sim 10^{-2}$ ($10^{-4}$) using $\sim 500,000$ physical qubits with the LCU (LT) encoding. With distance 31 the error drops to $\sim 10^{-5}$ ($10^{-7}$) with 750,000 physical qubits. These estimates use the Fowler-Gidney construction \cite{fowler_low_2019} with CCZ2T factories \cite{gidney_efficient_2019}.

\section{Exact resource estimate for Trotterized open system dynamics}
\label{app:ex}

Here, we will start with exactly computing the error of implementing $e^{\delta\mathcal{L}}$ with a single Lindblad generator $L_i$ following Theorem III.1 of \cite{chen_quantum_2023}, illustrated in Fig. \ref{fig:evo}. We then use a simple Trotter formula to combine them for different $L_i$. We note that using Corollary III.1 of \cite{chen_quantum_2023} to combine them probabilistically could, in principle, improve the error and T-counts. We do not pursue this as the Trotter approach is already suboptimal for all but the shortest times.

\subsection{Error of Fig. (\ref{fig:evo})}

Let the block encoding circuit $U$ acting on the system and the ancilla register $B$ encode only one Lindbladian $L = \langle 0|_B U |0\rangle_B$:
\begin{equation}
   U = \left(
\begin{array}{c|ccccccccc}
L & ~ & ~& ~& R &~ &~ & ~&~ &~\\
\hline
~ & ~& ~& ~ & ~ & ~ & ~& ~ &~& ~ \\
~ & ~& ~& ~ & ~ & ~ & ~& ~&~ & ~ \\
~ & ~ & ~& ~ & ~ & ~ &  ~ &~ & ~ & ~\\
M & ~& ~ & ~& D & ~ & ~ &~ &~ & ~ \\

~ & ~& ~ & ~& ~ & ~ & ~ &~ &~ & ~ \\
~ & ~ & ~& ~ & ~ & ~ &  ~ &~ & ~& ~ \\
~ & ~& ~ & ~& ~ & ~ & ~ &~ &~ & ~ \\
~ & ~& ~ & ~& ~ & ~ & ~ &~ &~ & ~\\
\end{array}
\right)
\end{equation}

We will use $UU^\dag =U^\dag U = {\mathds{1}}$ to derive relations $L^\dag L +M^\dag M = {\mathds{1}}$ and $R^\dag L + D^\dag M =0$. This $U$ and its controlled version $CU$ are used in the operation $ \mathcal{U}$ depicted in Fig. \ref{fig:evo}, acting on the system, the register $B$ and the two-qubit ancilla register $A$.
After a direct computation of tr$_{AB}(\mathcal{U}(P_{00,0_B} \otimes \rho) \mathcal{U}^\dag)$, we find:
\begin{align}
   \tilde{\mathcal{E}}(\rho) =\text{tr}_{AB} (\mathcal{U}(P_{00,0_B} \otimes \rho) \mathcal{U}^\dag) =\\= \rho + \delta L\rho L^\dag +(\sqrt{1-\delta} -1)\{L^\dag L,\rho\} +\\ + (\sqrt{1-\delta} -1)^2 (L^\dag L\rho L^\dag L + R^\dag L\rho L^\dag R)
\end{align}
This has to be compared with the target evolution:
\begin{equation}
    \mathcal{E}(\rho) = \text{exp}(\delta (L \bullet L^\dag - \frac{1}{2}\{L^\dag L,\bullet\}))\rho
\end{equation}
For comparison, we will need the Taylor theorem for the superoperator $\mathcal{E}(\rho)$. First, consider a scalar function $e^x$. The following is true for some $u \in[0,1]$:
\begin{equation}
    e^x = 1+ x +\frac{x^2}{2}e^{ux}
\end{equation}
Since all the series for exponentials converge, we can now formally plug in $\hat{x} =\delta (L \bullet L^\dag - \frac{1}{2}\{L^\dag L,\bullet\}$ and apply both sides to $\rho$:
\begin{equation}
     \mathcal{E}(\rho) = \rho + \delta (L \rho L^\dag - \frac{1}{2}\{L^\dag L,\rho\}) +  \frac{\hat{x}^2}{2}(\mathcal{E}^u(\rho))
\end{equation}
The difference becomes:
\begin{align}
   \tilde{\mathcal{E}}(\rho) - \mathcal{E}(\rho)  =\\= (\sqrt{1-\delta} -1+ \frac{\delta}{2})\{L^\dag L,\rho\} -\frac{\hat{x}^2}{2}(\mathcal{E}^u(\rho)) +\\ + (\sqrt{1-\delta} -1)^2 (L^\dag L\rho L^\dag L + R^\dag L\rho L^\dag R)
\end{align}
Let us first work with the trace norm $\|X\|_1$ of the above difference and the Holder inequality $\|XY\|_1 \leq \|X\| \|Y\|_1$ where $\|X\|$ is the operator norm. Also note that since $\mathcal{E}$ is a contractive map and $\rho\geq 0$ is trace 1, $\|\mathcal{E}^u(\rho)\|_1 =1$. Since $U$ is unitary, $\|L\| \leq 1, ~ \|R\|\leq 1$. Finally, $\|\hat{x}(\rho)\|_1 = 2\delta$. Putting it all together, we get: 
\begin{align}
  \| \tilde{\mathcal{E}}(\rho) - \mathcal{E}(\rho) \|_1 \leq\\\leq 2|\sqrt{1-\delta} -1+ \frac{\delta}{2}|+2\delta^2 +\\ + 2(\sqrt{1-\delta} -1)^2 
\end{align}
For now $\delta$ was constrained to $[0,1)$. The values at $1$ are
\begin{equation}
    2|\sqrt{1-\delta} -1+ \frac{\delta}{2}|_{\delta =1} =1, \quad (\sqrt{1-\delta} -1)^2_{\delta =1} =1 
\end{equation}
It can be checked that for $\delta \in [0,1]$:
\begin{equation}
    2|\sqrt{1-\delta} -1+ \frac{\delta}{2}| \leq \delta^2, \quad (\sqrt{1-\delta} -1)^2 \leq \delta^2 
\end{equation}
Restricting $\delta$ to smaller values is unlikely to lead to substantially better bounds. The final expression for error becomes:
\begin{equation}
    \| \tilde{\mathcal{E}}(\rho) - \mathcal{E}(\rho) \|_1 \leq 5\delta^2 \label{eq:beLerr}
\end{equation}

\paragraph{Note on the diamond norm}
The diamond norm of the channel $\mathcal{E}$ is defined as follows. First, define the induced superoperator norm as:
\begin{equation}
    \|\mathcal{E}\|_1 = \text{max}_X(\|\mathcal{E}(X)\|_1 ~ : ~ \|X\|_1 \leq 1 )
\end{equation}
Then the diamond norm is defined as a supremum over all possible larger systems where the channel $\mathcal{E}$ is applied to a subsystem:
\begin{align}
\left\Vert \mathcal{E} \right\Vert_{\diamond}^{} := \sup_{k \geq 1} \left\Vert \mathcal{E} \otimes \mathbb{I}_{k} \right\Vert_{1}^{},
\end{align}
where \(\mathbb{I}_{k}\) is the identity on a \(k\)-dimensional Hilbert space. It is well-known that the supremum is achieved for \(k = \mathrm{dim}(\mathcal{H})\), the dimension of the original Hilbert space on which the quantum channel acts \cite{wildeClassicalQuantumShannon2016}. Therefore, the diamond norm can be equivalently thought of as,
\begin{align}
\left\Vert \mathcal{E} \right\Vert_{\diamond}^{} = \max\limits_{X, \left\Vert X \right\Vert_{1}^{} \leq 1} \left\Vert \mathcal{E} \otimes \mathbb{I}_{\mathrm{dim}(\mathcal{H})} \right\Vert_{1}^{},
\end{align}
where \(\mathrm{dim}(\mathcal{H})\) is the dimension of the original Hilbert space on which \(\mathcal{E}\) acts.

The channel $\mathcal{E} \otimes \mathbb{I}_{k}$ applied to a general matrix $\rho_k = \sum_{ij}\rho_{ij} \otimes |i\rangle \langle j|$ where $i,j \in 1\dots k$ results in \begin{equation}
    \mathcal{E} \otimes \mathbb{I}_{k} (\rho_k) = \sum_{ij}\mathcal{E}(\rho_{ij}) \otimes |i\rangle \langle j|
\end{equation}
Above, we only proved the error difference for diagonal matrices of the form $\sum_i {\rho}_{ii} |i\rangle \langle i|$. Generally, the presence of off-diagonal terms can introduce additional factors of dimension into the norm. Following \cite{chen_quantum_2023}, we note that using the Lindblad operators $L \otimes  {\mathds{1}}_{k}$, the entire proof may be repeated. Since the dimension of $\rho$ itself never appeared in the computation of Eq. (\ref{eq:beLerr}), only the fact that $\|\rho\|_1 =1$, we can establish that for any $k$, and thus for diamond norm:
\begin{equation}
    \| \tilde{\mathcal{E}}(\rho) - \mathcal{E}(\rho) \|_\diamond \leq 5\delta^2
\end{equation}
\subsection{Error of Trotter}
The same Taylor theorem can be used to estimate the prefactors in the errors for the non-commuting $L_i,  H_j$:
\begin{align}
    \mathcal{E}_i(\rho) = \text{exp}(\delta (L_i \bullet L_i^\dag - \frac{1}{2}\{L_i^\dag L_i,\bullet\}))\rho  \\
    \mathcal{U}_j(\rho) = \text{exp}(-i\delta [H_j, \bullet] )\rho\\
    \mathcal{E}(\rho) = \text{exp}(-i\sum_j\delta[H_j, \bullet] + \\ +\delta \sum_i(L_i \bullet L_i^\dag - \frac{1}{2}\{L_i^\dag L_i,\bullet\}) )\rho 
\end{align}
The error is bounded as:
\begin{equation}
    \|\mathcal{E} - \prod_i\mathcal{E}_i\prod_j\mathcal{U}_j\|_1 =O(\delta^2)
\end{equation}
Here we assumed $\|H_i\|\leq 1$. Generally there's a scale factor between $\|H_i\|$ and $\|L_i\|$ which will enter the estimates. Like above, we can repeat the construction for $L_i \otimes  {\mathds{1}}_{k}$ and $H_i \otimes  {\mathds{1}}_{k}$ to establish the same result for
 the diamond norm:
 \begin{equation}
      \|\mathcal{E} - \mathcal{E}_1\mathcal{E}_2\mathcal{U}\|_\diamond =O(\delta^2)
 \end{equation}
 To prove the prefactors in the Trotter result, the specific form of the generator in the exponent does not matter, only the appropriate norm. Define:
 \begin{align}
     \mathcal{E}_i =\text{exp}(\delta \cdot\mathcal{L}_i(\rho)), \quad \mathcal{L}_i =(L_1 \bullet L_1^\dag - \frac{1}{2}\{L_1^\dag L_1,\bullet\} \\
     \mathcal{U}_i = \text{exp}(\delta \cdot\mathcal{G}_i )\rho \quad \mathcal{G}_i =-i [H_i, \bullet]
 \end{align}
 Consider the diamond norm $\|\mathcal{L}_i\|_\diamond, ~\| \mathcal{G}_i\|_\diamond$.  We need to take the maximum over dimension-$k$ extensions $L \otimes  {\mathds{1}}_{k}$ and $H \otimes  {\mathds{1}}_{k}$ of the trace norm:
 \begin{align}
    \|\mathcal{L}_i\otimes  {\mathds{1}}_{k}\|_1 \leq 2\|L_i \otimes  {\mathds{1}}_{k}\| = 2\|L_i\| \\\|\mathcal{G}_j\otimes  {\mathds{1}}_{k}\|_1 \leq  2\|H_j \otimes  {\mathds{1}}_{k}\| =  2\|H_j\|
 \end{align}
 Therefore, the diamond norm is bounded as well: 
  \begin{align}
    \|\mathcal{L}_i\|_\diamond \leq2\|L_i\| \\\|\mathcal{G}_j\|_\diamond  \leq    2\|H_j\|
 \end{align}
 Let us consider the case where $H =\sum_j H_j$ and the terms within each $H_j$ commute between themselves. The terms $L_j$ are joined into $J$ commuting groups, each containing $m$ such terms. With the new definition of one $\mathcal{L}_j$ per group, we get
 \begin{equation}
     \|\mathcal{L}_i\|_\diamond \leq2m\|L_i\|
 \end{equation}
 We are now ready to use the Taylor theorem:
 \begin{align}
      \mathcal{E}_i = 1 + \delta \mathcal{L}_i + \frac12 \delta^2 \mathcal{L}_i^2\mathcal{E}_i^u \\ 
      \mathcal{U}_j = 1 + \delta \mathcal{G}_j + \frac12 \delta^2 \mathcal{G}_j^2\mathcal{U}_j^u \\
      \mathcal{E} = 1 + \delta (\sum_i\mathcal{L}_i + \sum_j \mathcal{G}_j) +\\+ \frac12 \delta^2 (\sum_i\mathcal{L}_i + \sum_j \mathcal{G}_j)^2\mathcal{E}^u
 \end{align}
 The difference is:
 \begin{align}
    \mathcal{E} - \prod_i  \mathcal{E}_i\prod_j\mathcal{U}_j  = \\ = \frac12 \delta^2 (\sum_i\mathcal{L}_i + \sum_j \mathcal{G}_j)^2\mathcal{E}^u -R
 \end{align}
The remainder $R$ can be computed as follows. First, we unify notation so that $\mathcal{U}$ are also denoted as $\mathcal{E}$, and the index $i$ spans both. Then we define $p_k$ as partial products:
\begin{equation}
    p_k = \prod_{i\geq k} \mathcal{E}_i
\end{equation}
Now we denote the $\sim\delta$ and $\sim \delta^2$ terms as $x_i$ and $r_i$:
\begin{equation}
    \mathcal{E}_i = 1 +  x_i + r_i
\end{equation}
We find the following identities:
\begin{align}
    p_k -1 = \sum_{i\geq k}(x_i+r_i) p_{i+1} \\
  R =  \prod_i  \mathcal{E}_i - 1 - \sum_i x_i = \\ =\sum_i x_i (p_{i+1}-1) +r_i p_{i+1} = \\  =\sum_i\left( x_i \sum_{j\geq i+1}(x_j+r_j) p_{j+1} +r_i p_{i+1} \right)
\end{align}
The norms $\|p_j\|_\diamond \leq 1$ by construction. The remaining norm can be upper bounded as:
\begin{align}
    \|R\|_\diamond \leq \sum_i \|x_i\|_\diamond  \sum_j ( \|x_j\|_\diamond+ \|r_j\|_\diamond) +  \sum_i\|r_i\|_\diamond \\ \leq  \delta ( \sum_i\|\mathcal{L}_i\|_\diamond  + \sum_j \|\mathcal{G}_j\|_\diamond)  \times \\ \times ( \sum_i\delta\|\mathcal{L}_i\|_\diamond  +\frac12 \delta^2\| \mathcal{L}_i\|_\diamond^2 +\\+ \sum_j \delta \|\mathcal{G}_j\|_\diamond +\frac12 \delta^2 \|\mathcal{G}_j\|_\diamond^2 ) +\\+ \sum_i\frac12 \delta^2\| \mathcal{L}_i\|_\diamond^2 + \sum_j \frac12 \delta^2 \|\mathcal{G}_j\|_\diamond^2 \leq \\ \leq  \delta^2 ( \sum_i 2m\|L_i\|  + \sum_j 2\|H_j\|)  \times \\ \times ( \sum_i2m\|L_i\|  +\frac12 \delta(2m\|L_i\|)^2 +\\+ \sum_j  2\|H_j\| +\frac12 \delta (2\|H_j\|)^2 ) +\\+ \sum_i\frac12 \delta^2(2m\|L_i\|)^2 + \sum_j \frac12 \delta^2 (2\|H_j\|)^2
\end{align}
The total bound is then:
\begin{align}
   \| \mathcal{E} - \prod_i  \mathcal{E}_i\prod_j\mathcal{U}_j \|_\diamond \leq \\ \leq \frac32 \delta^2 (\sum_i2m\|L_i\| + \sum_j 2\|H_j\|)^2 + \\+\delta^2\left(\sum_i\frac12 (2m\|L_i\|)^2 + \sum_j \frac12  (2\|H_j\|)^2\right) \times \\ \times (1 +\delta (\sum_i2m\|L_i\| + \sum_j 2\|H_j\|))
 \end{align}
 At many steps, the bound can be tighter, and we ignored all the possible cancellations between $R$ and the other term. However, Trotter is already suboptimal for any except the shortest times, which is why we do not attempt to improve this bound further here.

 We will apply it to the Ca$_3$Co$_2$O$_6$ simulation. The Hamiltonian has two terms, the z-component with the norm $\|H_z\| \leq 1.6 |J_1| n$ (we assume zero longitudinal field for this estimate) per site and the x-component $\|H_x\| \leq \frac{|J_1|}{300} n$, where $n = 2025$  is the total number of spins, and we take the unit of energy $|J_1| =1$. The strength of each $\|L_i\| \leq 1$. As explained in the block encoding discussion, it is possible to tile the lattice by cells of $18$ spins, so in each group of commuting Lindblad operators there are $2025/18\leq m$ of them. The range of index $i =1 \dots 18*39=702$ for different spin flip rates. Plugging it into the bound results in the following:
 \begin{align}
 \sum_i2m\|L_i\| \leq 157950, \\ \sum_j 2\|H_j\| \leq 6493.5 
 \\ \sum_i(2m\|L_i\|)^2 \leq 35538750, \\ \sum_j (2\|H_j\|)^2 \leq 41990582.25
 \\  
   \| \mathcal{E} - \prod_i  \mathcal{E}_i\prod_j\mathcal{U}_j \|_\diamond \leq \\ \leq \frac32 \delta^2 (164443.5)^2 + \\+38764666.125\delta^2 ( 1 +164443.5\delta)
 \end{align}
 Say we want $n\times$this error to be $\leq \epsilon=0.01$, and $n\delta =T =10^{11}$, the evolution time, same as what was used for the optimal estimates. The requirement is then:
 \begin{align}
    10^{11} (  \frac32 \delta (164443.5)^2 + \\+38764666.125\delta ( 1 +164443.5\delta) ) \leq 0.01
 \end{align}
For $\delta = 2.4\cdot 10^{-24}$ it is satisfied. This translates to the number of steps $n = T/\delta =4.2 \cdot 10^{34}$. 

The Hamiltonian steps can be performed using $15\cdot2025/2$ two-qubit arbitrary angle rotations for $H_z$ (which can be performed via 4 CNOTS and 1 single qubit rotation on an ancilla), and $2025$ single qubit arbitrary angle rotations for $H_x$. The total number of single qubit rotations is then $8.5\cdot2025$ per Trotter step in the Hamiltonian application. The depth comes from $15\cdot2$ layers of the two-qubit rotations of depth $4$ and a rotation each, and one depth $1$ layer for the single qubit rotation, for a total of $120$ and $31$ rotations.

\begin{figure}
\includegraphics[width=0.75\columnwidth]{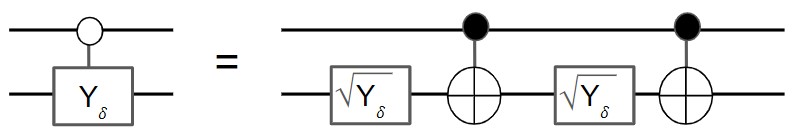}
\caption{A an implementation of a controlled rotation $Y_\delta= e^{-iY\text{arcsin}\sqrt{\delta}}$}
\label{fig:Crot}
\end{figure}

The total count for each $U_\mu$ and $CU_\mu$ was computed in the main text:
\begin{itemize}
    \item  $U_\mu$: $100$ T-gates, depth $418$
    \item $CU_\mu$: $108$ T-gates, depth $427$  
\end{itemize}
For the superoperator $\mathcal{E}_i$ containing $U_\mu$, $CU_\mu$ and a controlled rotation, we get:
\begin{itemize}
    \item T-count of $\mathcal{E}_i$: $208$
    \item depth except $R$: $847$
    \item $R$ count: $2$
\end{itemize}
There are $39\cdot2025$ superoperators $\mathcal{E}_i$, requiring $39\cdot18$ layers. This makes the count for the Trotter step (from both $\mathcal{E}_i$ and $H$):
\begin{itemize}
    \item T-count of $\mathcal{E}$: $16,426,800$
    \item $R$ count of $\mathcal{E}$: $175,163$
    \item depth without $R$: $594,714$
    \item $R$-depth: $1,435$
\end{itemize}
In \cite{kliuchnikov2023shorter} the average T-cost per single-qubit rotation is calculated to scale as:
\begin{equation}
   \frac{\text{\# T-gates}}{\text{rotation}} \sim 0.53\text{log}_2(1/\epsilon') + 4.86 
\end{equation}
Note that in the resource estimate paper \cite{beverland2022assessing} we used the constant contains a misprint. Fig. 3 in \cite{kliuchnikov2023shorter} shows that this scaling also works as an upper bound at large $\epsilon'$. 

The tool \cite{beverland2022assessing} works with the latter $\epsilon$ specified as input and allocates $\epsilon/3$ to each of the three sources of error, one of which is the rotation errors $\epsilon'$. The other two are surface code and distillation errors. We set the total of rotation and Trotter error $=0.02$ for this estimate.

As the algorithm has $n=4.2 \cdot 10^{34}$, the allowed error in each rotation is:
\begin{equation}
    \epsilon' = \frac{0.01}{175,163n} =1.4\cdot 10^{-42} 
\end{equation}
This leads to:
\begin{equation}
   \frac{\text{\# T-gates}}{\text{rotation}} \sim 0.53\text{log}_2(1/\epsilon') + 4.86  = 79
\end{equation}
Which results in a total counts:
\begin{itemize}
    \item T-count of $\mathcal{E}$: $30,264,677$
    \item T-count of the algorithm: $1.27\cdot 10^{42}$
\end{itemize}
We will assume that there is no parallelism within the rotation gates and use $2\cdot79$ as their depth. Parallelism will come from the invariance of the problem in translation. 
\begin{itemize}
    \item depth of $\mathcal{E}$: $821,444$
    \item depth of the algorithm: $3.45\cdot 10^{40}$
\end{itemize}

\end{document}